
\magnification 1200
\baselineskip=12pt
\hsize=15.3truecm \vsize=22 truecm \hoffset=.1truecm
\parskip=14 pt

\def\idty{{\leavevmode{\rm 1\mkern -5.4mu I}}}
\def\Ibb #1{ {\rm I\mkern -3.6mu#1}}
\def\Ird{{\hbox{\kern2pt\vbox{\hrule height0pt depth.4pt width5.7pt
    \hbox{\kern-1pt\sevensy\char"36\kern2pt\char"36} \vskip-.2pt
    \hrule height.4pt depth0pt width6pt}}}}
\def\Irs{{\hbox{\kern2pt\vbox{\hrule height0pt depth.34pt width5pt
       \hbox{\kern-1pt\fivesy\char"36\kern1.6pt\char"36} \vskip -.1pt
       \hrule height .34 pt depth 0pt width 5.1 pt}}}}
\def\Ir{{\mathchoice{\Ird}{\Ird}{\Irs}{\Irs} }}
\def\ibbt #1{\leavevmode\hbox{\kern.3em\vrule
     height 1.5ex depth -.1ex width .2pt\kern-.3em\rm#1}}
\def\ibbs#1{\hbox{\kern.25em\vrule
     height 1ex depth -.1ex width .2pt
                   \kern-.25em$\scriptstyle\rm#1$}}
\def\ibbss#1{\hbox{\kern.22em\vrule
     height .7ex depth -.1ex width .2pt
                   \kern-.22em$\scriptscriptstyle\rm#1$}}
\def\ibb#1{{\mathchoice{\ibbt #1}{\ibbt #1}{\ibbs #1}{\ibbss #1}}}
\def\Nl{{\Ibb N}} \def\Cx {{\ibb C}} \def\Rl {{\Ibb R}}
\def\lessblank{\parskip=5pt \abovedisplayskip=2pt
          \belowdisplayskip=2pt }
\def\eproclaim{\par\endgroup\vskip0pt plus100pt\noindent}
\def\proof#1{\par\noindent {\bf Proof #1}\          
         \begingroup\lessblank\parindent=0pt}
\def\QED {\hfill\endgroup\break
     \line{\hfill{\vrule height 1.8ex width 1.8ex }\quad}
      \vskip 0pt plus100pt}
\outer\def\example#1/#2/ {\vskip0pt plus50pt \par\noindent
     {\bf\dpcl#1/#2/ Example.\ }\begingroup
      \interlinepenalty=250\lessblank}
\def\endexample{\endgroup\hfill $\bigtriangleup$}
\def\authorbox#1=#2\\#3\email#4\\{{\let\\\cr\hbox{\vtop{
             \halign{\hfil##\hfil\cr
                  \bf#1$^{#2}$\cr
                    \global\edef\footstuff{\footstuff
                    \leftline{$^{#2}$\quad Email: \tt#4}}\cr
                  \noalign{\smallskip}
                  #3\crcr}}}}}
\def\footstuff{\vfill\hrule width5cm \vskip4pt}


\def\Set#1#2#3{\ifx#2:#1\lbrace{#3}#1\rbrace\else
                     {#1\lbrace{#2}#1\vert{#3}#1\rbrace}\fi}
\def\Abs#1#2{#1\vert#2#1\vert}       
\def\Norm#1#2{#1\Vert#2#1\Vert}      %
\def\abs #1{{\left\vert#1\right\vert}}
\def\bra #1>{\langle #1\rangle}
\def\bracks #1{\lbrack #1\rbrack}
\def\bs{{\backslash}}                                  

\def\id{\mathop{\rm id}\nolimits}
\def\ker{{\rm ker}}
\def\ket #1>{\mid#1\rangle}
\def\norm #1{\left\Vert #1\right\Vert}
\def\cbnorm #1{\norm{{#1}}_{\rm cb}}
\def\set #1{\left\lbrace#1\right\rbrace}
\def\stack#1#2{{\buildrel #1 \over #2}}           
\def\th{\hbox{${}^{{\rm th}}$}\ }  

\def\trnm#1{\vert\mkern-2mu\vert\mkern-2mu\vert      
           #1\vert\mkern-2mu\vert\mkern-2mu\vert}

\def\trnmcb#1{\trnm{#1}_{\rm cb}}
\def\svert{\overline{\vert\mkern-2mu\vert\mkern-2mu\vert}}
\def\strnm#1{\svert{#1}\svert}   

\font\bigsym=cmsy10 scaled \magstep4
\def\bigtimes{\mathop{\lower2pt\hbox{\bigsym\char2}}}
\def\\{\hfill\break}
\def\ie{i.e., }         
\def\eg{e.g., }         

\def\cf#1{{{\cal C}\left({#1}\right)}}  
\def\rar{{\;\rightarrow\;}}           
\def\ep{\epsilon}                     
\def\H{{\cal H}}                      
\def\K{{\cal K}}                      
\def\B{{\cal B}}                      
\def\At{\widetilde{\cal A}}           
\def\A{{\cal A}}                      
\def\C{{\cal C}}                      
\def\M{{\cal M}}                      
\def\de{{\delta}}                     
\def\La{{\cal L}}                     
\def\eth{{\hat{\eta}}}                
\def\condex{{\Ibb E}}
\def\Subc{S}   
\def\dd(#1)#2{d_{\scriptstyle #1}\ifx\cdot#2\else#2\fi#2} 
\def\dd(#1)#2{d(#1;#2)}                                   
\def\fin{_{\rm fi\mkern-2mu n}}       
\def\loc{_{\rm loc}}
\def\srp{\sigma'}    
\def\diag{{\bf M}} 

\catcode`@=11
\def\ifundefined#1{\expandafter\ifx\csname
                        \expandafter\eat\string#1\endcsname\relax}
\def\atdef#1{\expandafter\def\csname #1\endcsname}
\def\atedef#1{\expandafter\edef\csname #1\endcsname}
\def\atname#1{\csname #1\endcsname}
\def\ifempty#1{\ifx\@mp#1\@mp}
\def\ifatundef#1#2#3{\expandafter\ifx\csname#1\endcsname\relax
                                  #2\else#3\fi}
\def\eat#1{}
\newcount\refno \refno=1
\def\labref #1 #2 #3\par{\atdef{R@#2}{#1}}
\def\lstref #1 #2 #3\par{\atedef{R@#2}{\number\refno}
                              \advance\refno by1}
\def\txtref #1 #2 #3\par{\atdef{R@#2}{\number\refno
      \global\atedef{R@#2}{\number\refno}\global\advance\refno by1}}
\def\doref  #1 #2 #3\par{{\refno=0
     \vbox {\everyref \item {\reflistitem{\atname{R@#2}}}
            {\d@more#3\more\@ut\par}\par}}\refskip }
\def\d@more #1\more#2\par
   {{#1\more}\ifx\@ut#2\else\d@more#2\par\fi}
\def\@cite #1,#2\@ver
   {\eachcite{#1}\ifx\@ut#2\else,\hskip0pt\@cite#2\@ver\fi}
\def\cite#1{\citeform{\@cite#1,\@ut\@ver}}
\def\eachcite#1{\ifatundef{R@#1}{{\tt#1??}}{\atname{R@#1}}}
\def\defonereftag#1=#2,{\atdef{R@#2}{#1}}
\def\defreftags#1, {\ifx\relax#1\relax \let\next\relax \else
           \expandafter\defonereftag#1,\let\next\defreftags\fi\next }
\def\@utfirst #1,#2\@ver
   {\author#1,\ifx\@ut#2\afteraut\else\@utsecond#2\@ver\fi}
\def\@utsecond #1,#2\@ver
   {\ifx\@ut#2\andone\author#1,\afterauts\else
      ,\author#1,\@utmore#2\@ver\fi}
\def\@utmore #1,#2\@ver
   {\ifx\@ut#2\and\author#1,\afterauts\else
      ,\author#1,\@utmore#2\@ver\fi}
\def\authors#1{\@utfirst#1,\@ut\@ver}
\catcode`@=12
\let\REF\labref
\def\citeform#1{{\bf\lbrack#1\rbrack}}  
\let\reflistitem\citeform               
\let\everyref\relax                     
\let\more\relax                         
\def\refskip{\vskip 10pt plus 2pt}      
\def\colbr{\hskip.3em plus.3em\penalty-100}  
\def\combr{\hskip.3em plus4em\penalty-100}   
\def\refsecpars{\emergencystretch=50 pt      
                 \hyphenpenalty=100}
\def\Bref#1 "#2"#3\more{\authors{#1}:\colbr {\it #2},\combr #3\more}
\def\Gref#1 "#2"#3\more{\authors{#1}\ifempty{#2}\else:\colbr``#2''\fi
                        ,\combr#3\more}
\def\Jref#1 "#2"#3\more{\authors{#1}:\colbr``#2'',\combr \Jn#3\more}
\def\inPr#1 "#2"#3\more{in: \authors{\eds#1}:\colbr
                          ``{\it #2}'',\combr #3\more}
\def\Jn #1 @#2(#3)#4\more{\hbox{\it#1}\ {\bf#2}(#3)#4\more}
\def\author#1. #2,{\hbox{#1.~#2}}            
\def\sameauthor#1{\leavevmode$\underline{\hbox to 25pt{}}$}
\def\and{, and}   \def\andone{ and}          
\def\noinitial#1{\ignorespaces}
\let\afteraut\relax
\let\afterauts\relax
\def\etal{\def\afteraut{, et.al.}\let\afterauts\afteraut
           \let\and,}
\def\eds{\def\afteraut{(ed.)}\def\afterauts{(eds.)}}
\catcode`@=11
\newcount\eqNo \eqNo=0
\def\lasteq{\secNo.\number\eqNo}
\def\deq#1(#2){{\ifempty{#1}\global\advance\eqNo by1
       \edef\n@@{\lasteq}\else\edef\n@@{#1}\fi
       \ifempty{#2}\else\global\atedef{E@#2}{\n@@}\fi\n@@}}
\def\eq#1(#2){\edef\n@@{#1}\ifempty{#2}\else
       \ifatundef{E@#2}{\global\atedef{E@#2}{#1}}%
                       {\edef\n@@{\atname{E@#2}}}\fi
       {\rm($\n@@$)}}
\def\deqno#1(#2){\eqno(\deq#1(#2))}
\def\deqal#1(#2){(\deq#1(#2))}
\def\eqback#1{{(\advance\eqNo by -#1 \lasteq)}}

\def\eqgroup(#1){{\global\advance\eqNo by1
       \edef\n@@{\lasteq}\global\atedef{E@#1}{\n@@}}}
\outer\def\iproclaim#1/#2/#3. {\vskip0pt plus50pt \par\noindent
     {\bf\dpcl#1/#2/ #3.\ }\begingroup
       \interlinepenalty=250\lessblank\sl}
\newcount\pcNo  \pcNo=0
\def\lastpc{\number\pcNo} 
\def\dpcl#1/#2/{\ifempty{#1}\global\advance\pcNo by1
       \edef\n@@{\lastpc}\else\edef\n@@{#1}\fi
       \ifempty{#2}\else\global\atedef{P@#2}{\n@@}\fi\n@@}
\def\pcl#1/#2/{\edef\n@@{#1}%
       \ifempty{#2}\else
       \ifatundef{P@#2}{\global\atedef{P@#2}{#1}}%
                       {\edef\n@@{\atname{P@#2}}}\fi
       \n@@}
\def\Def#1/#2/{Definition~\pcl#1/#2/}
\def\Thm#1/#2/{Theorem~\pcl#1/#2/}
\def\Lem#1/#2/{Lemma~\pcl#1/#2/}
\def\Prp#1/#2/{Proposition~\pcl#1/#2/}
\def\Cor#1/#2/{Corollary~\pcl#1/#2/}
\def\Exa#1/#2/{Example~\pcl#1/#2/}
\font\sectfont=cmbx10 scaled \magstep2
\def\bgsecti@n #1. #2\e@h{\def\secNo{#1}\eqNo=0}
\def\bgssecti@n#1. #2\e@h{}
\def\secNo{00}
\def\lookahead#1#2{\vskip\z@ plus#1\penalty-250
  \vskip\z@ plus-#1\bigskip\vskip\parskip
  {#2}\nobreak\smallskip\noindent}
\def\secthead#1. #2\e@h{\leftline{\sectfont
                        \ifx\n@#1\n@\else#1.\ \fi#2}}
\def\bgsection#1. #2\par{\bgsecti@n#1. #2\e@h
        \lookahead{.3\vsize}{\secthead#1. #2\e@h}}
\def\bgssection#1. #2\par{\bgssecti@n#1. #2\e@h
        \lookahead{.3\vsize}{\leftline{\bf#1. #2}}}
\def\bgsections#1. #2\bgssection#3. #4\par{%
        \bgsecti@n#1. #2\e@h\bgssecti@n#3. #4\e@h
        \lookahead{.3\vsize}{\vtop{\secthead#1. #2\e@h\vskip10pt
                             \leftline{\bf#3. #4}}}}
\def\Acknow#1\par{\ifx\REF\doref
     \bgsection. Acknowledgements\par#1\refsecpars
     \bgsection. References\par\fi}
\catcode`@=12
\def\class#1 #2*{{#1},}
\overfullrule=0pt
\defreftags Bia=Biafore, BR=BraRo, DiV=Vincenzo, DS=Dunfo,
EHPP=qcrypt, FNW1=FCS, FNW2=FCP, FGSS=Zei2, GD=GD, GZ=Zei1,
LMS=Lebowitz, LTPH=Lent, Lig=Liggett, Mae=Krist, MS1=MSa, MS2=MSc,
Mai=Mainieri, MZ=Boguslav, Ma1=MATA, Ma2=MATB, Ma3=MATC, Ma4=MATD,
Pau=PAUc, Ric=PHD, Smi=SMI, Sti=STI, Tak=TAK, Wit=WIT, ,

{\nopagenumbers
\line{{\tt cond-mat/9504001}\hfill Preprint KUL-TF-95/10}
\vskip 1.0cm
\font\BF=cmbx10 scaled \magstep 3
{\BF \baselineskip=25pt
\centerline{Ergodicity of}
\centerline{Quantum Cellular Automata}
}
\vskip 2.0cm
\centerline{\authorbox
S. Richter=1            \\
Inst. Theor. Fysica     \\
Universiteit Leuven     \\
B--3001 Heverlee        \\
Belgium                 \\
\email susanne.richter@fys.kuleuven.ac.be\\
\hskip2truecm\authorbox
R.F. Werner=2           \\
FB Physik               \\
Universit\"at Osnabr\"uck   \\
D--49069 Osnabr\"uck   \\
Germany                 \\
\email  reinwer@dosuni1.rz.uni-osnabrueck.de\\
}
\vskip 1.0cm
{\baselineskip=12pt
\narrower\narrower\noindent
{\bf Abstract.}\
We define a class of dynamical maps on the quasi-local algebra of a
quantum spin system, which are quantum analogues of probabilistic
cellular automata. We develop criteria for such a system to be
ergodic, i.e., to possess a unique invariant state. Intuitively,
ergodicity obtains if the local transition operators exhibit
sufficiently large disorder.
The ergodicity criteria also imply bounds for the exponential decay
of correlations in the unique invariant state. The main technical
tool is a quantum version of oscillation norms, defined in the
classical case as the sum over all sites of the variations of an
observable with respect to local spin-flips.
\par}
\vskip20 pt
\noindent{\bf Math.Review Classification (1991):}
60K35, 82C10, 82C20, 81Q10, 60B12 \hfill\break
{\bf PACS Classification (1994):}
64.60.Cn, 75.10.Jm, 03.65.Db, 05.30.-d, 05.50.+q \hfill\break
\footstuff\eject}

\bgsection 1. Introduction

A Probabilistic Cellular Automaton (PCA) \cite{MSa,Lebowitz,GD}, or
interacting particle system \cite{Liggett}, can be regarded as an
infinite collection of cells or particles, where each cell or
particle can take on a finite number of states. The discrete time
evolution of such systems is determined by a statistical law
according to which, in any given configuration at time $t$, all
cells are simultaneously and independently updated to the
configuration at time $t+1$. One of the basic questions concerning
such systems is ergodicity, \ie the uniqueness of the stationary
probability measure on configurations. In this article we will
introduce a quantum analogue of this structure, and we will also
prove an analogue of well-known criteria by which ergodicity of a
PCA can be decided in terms of the local transition probabilities
\cite{Liggett,Lebowitz,MSa}.

Our definition of quantum cellular automata (QCA) is an abstraction
of a structure that arose in the project of generalizing the
construction of ``finitely correlated states'' on quantum spin
chains \cite{FCS,FCP} to two-dimensional systems \cite{PHD}. In this
context the ergodicity of the QCA is equivalent to the state on the
two-dimensional system to be uniquely determined by local data,
independently of boundary conditions. The term ``quantum cellular
automaton'' has been used previously by some other authors
\cite{Zei1,Zei2,Lent}. In the cases we are aware of, however, it is used
for a structure on the Hilbert space level, and not on the level of
observables. Thus in \cite{Zei1} the classical states at each site
are simply replaced by the values of the wave function at that site,
and the dynamics is just a discrete Schr\"odinger equation with
non-selfadjoint Hamiltonian, made non-linear by keeping the
normalization fixed. There is some interest in quantum cellular
automata also from the point of view of nanometer scale computers,
for which quantum effects are expected to be relevant
\cite{Mainieri,Biafore,Lent}. The evolution of the automata
considered in this paper is in general non-unitary, \ie pure states
may evolve into mixed states. This might be an interesting addition
to the structure of ``quantum computers'', as studied by a number of
authors recently (see \cite{Vincenzo}, and references cited there).

The main technical contribution of this article is the introduction
of a general class of ``oscillation norms'' on quantum lattice
systems. We believe this to be a useful tool of independent
interest, and therefore included proofs of the basic general
properties of such norms. They generalize a classical notion, which
also was a principal tool for the proof of the ergodicity criteria
in \cite{Lebowitz,MSa,MSc}. A special case of such a norm in the quantum
case was already used extensively by Matsui
\cite{MATA,MATB,MATC,MATD}. Among other things, he used it to
establish ergodicity criteria for the continuous time analogue of
QCAs. Another special case of oscillation norms, used for the same
purpose, is to be found in a recent preprint by Majewski and
Zegarlinski \cite{Boguslav}.

An important consideration in the study of transition operators on
composite quantum systems is that the norm of an operator acting on
observables may increase, if we consider the given system as a
subsystem of a larger one. It is therefore essential to consider
versions the basic operator properties, which are ``stabilized with
respect to system enlargement''. The stabilized versions of
positivity and boundedness (called ``complete'' positivity and
boundedness) are well-known, and we give a brief summary of these
with references in an Appendix. For the notion of ``boundedness in
oscillation norm'' the stabilized version is described in Section 4.
Much to our surprise, it turned out that in this case the necessity
of stabilization is not characteristic of the quantum case. Even in
the classical case, as soon as one has more than two states at each
site, the oscillation norm bound of a transition operator may
increase with the size of the environment, as we will show by an
explicit example.

The paper is organized as follows.
In Section 2 we establish our notations for quantum lattice systems,
and develop the definition of QCAs from the classical notion of
PCAs. Section 3.1 introduces oscillation norms in general
C*-algebras, and shows how contractivity of a transition operator in
such a norm entails ergodicity. Section 3.2 is devoted to the
construction of a canonical oscillation norm for a system composed
of many parts with given oscillation norms, such as the quasi-local
algebra. The key result, reducing estimates on the infinite systems
to an estimate of a local quantity, is shown in Section 4.1: if the
local transition operators contract with a certain rate, as measured
by the ``completely bounded oscillation norm'', then the same holds
for an infinite tensor product of such operators. We briefly
indicate in Section 4.2 how other plausible approaches to such
estimates fail on this account. The basic ergodicity criterion for
QCAs in Section 5.1 is a direct corollary of this estimate. In 5.2
we show in what sense the classical PCAs are covered by this
criterion. Finally, we show, in Section 5.3, that the same estimate
on local transition operators, which implies ergodicity by our
general criterion, also entails exponential decay of correlation
functions in the unique invariant state. A quick review of the
notions of complete positivity and complete boundedness, with
pointers to the literature, is given in the Appendix.

\bgsection 2. Definition of Quantum Cellular Automata

In order to describe the notion of quantum cellular automata (QCA),
it is best to begin by restating the classical structure in an
algebraic language more suitable for generalization to the quantum
case. The underlying lattice, or set of cells, will be denoted by
$\La$. To each cell $x\in\La$ we associate an observable algebra
$\A^x$, which in the case of a classical system is simply the
algebra $\A^x=\cf{\Omega^x}$ of continuous complex valued functions
on the set $\Omega^x$ of configurations of each cell. Finite
subsystems, associated with finite subsets $\Lambda\subset\La$ are
described by the tensor product
$$ \A^\Lambda=\bigotimes_{x\in\Lambda} \A^x
    =\cf{\bigtimes_{x\in\Lambda}\Omega^x}
\quad.\deqno(fintensor)$$
This formula is also used for infinite subsystems, and in
particular, for the observable algebra $\A^\La$ of the whole system.
The product on the right hand side is then the infinite Cartesian
product of compact topological spaces. Even if each cell has only
finitely many configurations, and hence ``continuity'' for functions
on each $\Omega^x$ is a vacuous condition, continuity of the
observables $f\in\A^\La$ is the non-trivial requirement that $f$ may
be uniformly approximated by observables depending only on finitely
many cells.

The PCA dynamics is determined by the transition functions
$$p_x(\omega,\sigma)
    =\condex\Set\Big{\omega_{x,t+1}\in\sigma}{\omega_t=\omega}
\quad,$$
for $x\in\La$, $\omega\in\Omega$, and measurable
$\sigma\subset\Omega^x$. This is a probability measure on $\Omega^x$
in its second argument, depending, in principle on the entire
previous configuration $\omega\in\Omega$. In algebraic language this
becomes an operator
$$\eqalign{
               P_x&:\A^x\to\A^\La   \cr
   (P_x f)(\omega)&=\int p_x(\omega,d\omega_x)\ f(\omega_x)
\quad.}\deqno(PCA1site)$$
The assumption that $P_xf$ is a continuous function is called the
Feller property of the PCA \cite{MATD,Liggett}. It means
intuitively that the updating of one cell does not depend too
sensitively on infinitely many other cells. It is automatically
satisfied for finite range interactions, \ie when
$P_x(\A^x)\subset\A^{\Lambda(x)}\subset\A^\La$, for some finite set
$\Lambda(x)$.

Independent updating of different cells, the basic property of PCAs,
means in more formal language that the distribution of
$\omega_{t+1}$, given $\omega_t$, is the product measure formed out
of the measures $p_x(\omega_t,\cdot)$. Equivalently, the transition
operator $P:\A^\La\to\A^\La$ is defined as
$$ P\left(\bigotimes_{x\in\Lambda}f^x\right)
        =\prod_{x\in\Lambda} P_x(f^x)
\quad,\deqno(indup)$$
where $f^x\in\A^x$, $\Lambda\subset\La$ is an arbitrary finite set,
and the product on the right hand side is the pointwise product of
functions in the algebra $\A^\La=\cf\Omega$. Since the tensor
products on the left hand side of \eq(indup) span the C*-algebra
$\cf\Omega$, this equation determines $P$ uniquely. This concludes
our brief description of PCA dynamics.

Some of the above is easily translated into the {\it quantum
setting\/}: the main change is that now all observable algebras may
be non-commutative (rather than commutative) C*-algebras with
identity. For the observables algebra of a single cell one typically
chooses the algebra $\M_n$ of $n\times n$-matrices with $n<\infty$.
The simplest example is a Heisenberg spin-$1/2$ system, for which
$\A^x=\M_2$ for every $x\in\La$. The observable algebra of a
composite quantum system is defined as the closure of the algebraic
tensor product in a suitable C*-norm. In contrast to the classical
case there may be several such norms, in which case we always take
the ``minimal'' C*-norm \cite{TAK}. For the product of finite
dimensional matrix algebras, in which  we will mostly be interested,
all C*-tensor norms coincide anyhow. We continue to use the notation
$\A^\Lambda$ for the observable algebras \eq(fintensor) of finite
regions. For $\Lambda_1\subset\Lambda_2$, there is a natural
inclusion $\A^{\Lambda_1}\subset\A^{\Lambda_2}$, by tensoring each
element of $\A^{\Lambda_1}$ with the identity in
$\A^{\Lambda_2\setminus\Lambda_1}$. The {\it infinite tensor
product} defining the observable algebra of an infinite (sub-)system
is always defined as the C*-inductive limit \cite{BraRo} of the
finite tensor products with respect to these inclusions. As in the
classical case, this simply means that all observables can be
approximated in norm by finitely localized ones. A {\it transition
operator} such as $P_x$ can be characterized as a positive operator
(\ie an operator taking positive elements into positive elements),
mapping the identity into the identity. Moreover, we will also
assume that these properties persist if we consider $\A^x$ and
$\A^\La$ as subsystems of a larger system, \ie we require $P_x$ to
be completely positive (see the Appendix for definitions). A
probability measure is replaced in the algebraic framework by the
expectation value functional, or ``{\it state}'' it induces. Thus
states are linear functionals on the observable algebra, which take
positive values on positive elements, and the value $1$ on the
identity.

The key problem for the quantum generalization of PCAs is the
positivity of the right hand side of equation \eq(indup): in the
non-commutative context a product of positive elements is
practically never positive. In fact, restricting to the case of just
two factors ($\Lambda=\set{x,y}$ with $f^x,f^y\geq0$), we find that
a necessary and sufficient condition for the positivity of $P$ is
that the ranges of the operators $P_x$ commute in $\A^\La$ (see
Proposition IV.4.23(ii) in \cite{TAK}).

We therefore have to choose our definition in such a way that the
commuting range condition holds automatically. The following is one
way of doing this. It is perhaps not the most general possibility,
but it covers the cellular automata which came up naturally in our
construction of states on two-dimensional spin systems. The idea is
to subdivide each cell into subcells, such that the images of
different $P_x$ are contained in different tensor factors with
respect to the refined tensor decomposition of $\A^\La$. For
notational convenience we state the definition only in the case that
the observable algebras and transition operators of different cells
are all isomorphic.

\iproclaim/QCA/ Definition.
A {\bf quantum cellular automaton} (QCA) is given by the following
objects:
\item{(1)}
a countable lattice $\La$,
\item{(2)}
a set $\Subc$ of ``subcell types'', and a C*-algebra $\B^s$ with
unit, for each type $s\in\Subc$,
\item{(3)}
for each $s\in\Subc$, an injective ``propagation map''
$\dd(s){\cdot}:\La\to\La$, and
\item{(4)}
a completely positive unit preserving operator
$P_1:\bigotimes_{s\in\Subc}\B^s\to\bigotimes_{s\in\Subc}\B^s$.

\noindent
The following objects are defined in terms of the above:
\item{(5)}
at each site $x\in\La$ the algebra
$\A^x=\bigotimes_{s\in\Subc}\B^{x,s}$, where $\B^{x,s}$ is an
isomorphic copy of $\B^s$,
\item{(6)}
the quasi-local algebra $\A^\La=\bigotimes_{x\in\La}\A^x$,
\item{(7)}
the transition operators
$$   P_x:\A^x=\bigotimes_{s\in\Subc}\B^{x,s}
            \ \longrightarrow
            \ \bigotimes_{s\in\Subc}\B^{\dd(s)x,s}\quad
                \subset\A^\La
\quad,$$
where $P_x$ is defined from $P_1$ by identifying the tensor factor
$\B^s$ in the range of $P_1$ with
$\B^{\dd(s)x,s}\subset\A^{\dd(s)x}$, and
\item{(8)}
the total transition operator $P:\A^\La\to\A^\La$, defined by
equation \eq(indup).
\eproclaim

To see that the operator in step (8) is well-defined, observe that
the subcells are labelled by $\La\times\Subc$, and that the sets
$R^x=\Set{}{(\dd(s)x,s)}{s\in\Subc}\subset \La\times\Subc$, which describe
the range of $P_x$ are disjoint. Then, by Proposition IV.4.23(i) in
\cite{TAK}, $P$ is completely positive. Since $P\idty=\idty$, this
implies that $P$ is norm continuous, and consequently has a unique
extension by continuity to the whole quasi-local algebra $\A^\La$.

Of course, when we think of a lattice, the injective maps
$\dd(s){\cdot}$ will typically be lattice translations. If we choose
all $\dd(s){\cdot}$ to be the identity, we obtain a system of
non-interacting cells $\A^x$. Note that the subcell decomposition is
only relevant in the range of $P_x$, not in the domain. Thus in each
step the subcell decomposition of the previous step is obliterated.
In particular, the second iterate $P^2$ of a QCA cannot be written
in the same form: the algebras $P^2(\A^x)$ do not commute with each
other. This does not contradict the necessity of the commuting range
condition explained above, because the product form of $P^2$ is also
lost. Note that this is not an artefact of our quantum
generalization: even in the classical case the second generation
updates of a PCA are no longer independent, hence they no longer
satisfy the definition of a PCA.

The adjoint operator of $P$ takes states into states, and we will
usually denote its action by $\omega\mapsto \omega\circ P$. It is
easy to see (\eg using the Markov-Kakutani Theorem, Theorem~V.10.6
in \cite{Dunfo}) that any QCA has an {\it invariant state}, \ie a
state $\rho$ such that $\rho\circ P=\rho$. A QCA is called {\it
ergodic}, if there is only one invariant state for $P$. The main
problem addressed in this paper is to find sufficient criteria for
ergodicity in terms of the given local data $P_1$ and
$\dd(s){\cdot}$. We are also interested in stronger versions of
ergodicity, \eg the property that $P^n$ contracts in norm to the
invariant state, \ie
$$ \lim_{n\to\infty}\norm{P^n(A) - \rho(A)\idty}=0
\quad,$$
for all $A\in\A^\La$. A further closely related problem is to
estimate the decay of correlations in the invariant state.

To see what is involved, it is good to look at the most trivial
example: a non-interacting particle system. Then we have only one
type of subcells, and $\dd(1){\cdot}=\id_\La$. $P$ is simply the
infinite tensor product of copies $P_x$ of a fixed operator $P_1$,
acting on isomorphic finite dimensional algebras $\A^x$. It is
obvious that the restriction of an invariant state for $P$ to a
single site is invariant for $P_1$, and, conversely, any product
state formed out of invariant one-site states will be invariant for
$P$ (the latter construction need not be exhaustive). Hence $P$ is
ergodic if and only if $P_1$ is ergodic. It is plausible that
contractivity properties should also carry over. Assume, for
example, that
$$  \norm{P_1^n(A)-\rho_1(A)\idty}
       \leq\ep^n\norm{A-\rho_1(A)\idty}
\quad,$$
for all $A\in\A^x$, and $\rho_1$ the unique invariant state of
$P_1$. Does this imply a similar bound for $P$? A direct estimate
gives indeed a similar bound, but with $\ep^n$ multiplied by the
number of sites (see Section 4.2). Hence this approach is not
feasible on an infinite lattice. What one needs is a norm such that
a contractivity estimate for a tensor product of completely positive
operators $P_x$ is not worse than the maximum of the estimates for
the factors. This is precisely the role of the oscillation norms
used in the classical results of \cite{Lebowitz}. Their quantum
analogue will be studied in the following two sections. We will then
return to QCAs in Section 5.

\bgsections 3.   Oscillation norms on C*-algebras
\bgssection 3.1. Definition and basic properties

In this section we want to generalize the notion of oscillation norm
to the non-commutative setting. The basic idea remains the same: we
consider some operations $\de_\alpha$, which annihilate constants,
\ie $\de_\alpha(\idty)=0$. In the classical case these operations
measure the effect of the spin flips at different sites, \ie
$$ \bigl(\de_\alpha f\bigr)(\sigma_\alpha,\sigma_{\rm not\,\alpha})
    ={1\over2}\bigl(f(-\sigma_\alpha,\sigma_{\rm not\,\alpha})
                   -f(\sigma_\alpha,\sigma_{\rm not\,\alpha})\bigr)
\quad,\deqno(Isingosno)$$
where $\sigma_{\rm not\,\alpha}$ stands for all spin variables at
sites other than $\alpha$.
Then we can say that an observable $f$ is nearly constant, if the
``oscillation norm'' $\sum_\alpha\norm{\de_\alpha(f)}$ is small for
all $\alpha$.

For classical systems with more than two states per cell, as well as
for any quantum system, there are many ways of ``flipping'' a single
cell. There are several proposals in the literature how to take this
into account. However, all proposals agree that the ``total
oscillation norm'' of a lattice system should be the sum of the
oscillations of each cell. Moreover, we will see below that the total
oscillation norms in these different approaches are equivalent,
whenever the local cells are described by finite dimensional
algebras.
Perhaps the simplest proposal
\cite{Boguslav} is to define the oscillation of an observable $A$
localized in a single cell as
$$ \trnm A_0=\norm{A-\eta(A)\idty}
\quad,\deqno(bogus)$$
where $\eta$ is the normalized trace (or any other state) on the
cell algebra $\A^x$. Other approaches use a family
$\set{\de_\alpha}$ of operators for each cell, and one can consider
the ``sup-oscillation norm''
$$ \strnm A=\sup_\alpha\norm{\de_\alpha(A)}
\quad.\deqno(strnm)$$
It is also suggestive to define
$$ \trnm A_d=\sup_{\eta,\eta'}{\abs{\eta(A)-\eta'(A)}
                                \over d(\eta,\eta')}
\quad,\deqno (Krist)$$
where the supremum is over all pairs of states, and $d$ is some
metric on the state space. In the classical case one could restrict
the supremum to pure states, so that $\trnm A_d$ is just the
Lipshitz constant of $A$ with respect to the metric $d$
\cite{Krist}. Finally, one may use for the single cell precisely the
same form as for the total oscillation, namely a sum
$$\trnm{A}:=\sum_{\alpha}\Norm\big{\de_{\alpha}(A)}
\deqno(sumtrnm)$$
over ``elementary'' oscillations $\de_\alpha(A)$. This is the
approach used by Matsui \cite{MATA}. We will also adopt it, mainly
because it agrees best with the subcell structure of cellular
automata: the propagation maps will introduce a reshuffling of
subcells between different main cells, and this process preserves
oscillation norms only if the oscillation within each cell is
defined by the same mechanism as the total oscillation. This
approach also simplifies the presentation in the sense that we can
use the same results about oscillation norms for the single cells as
well as for the whole system.

\iproclaim/osno/Definition.
Let $\A$ be a $C^*$-algebra, $\idty\in\A$. Let $I$ be an index set
and $\Set\big{\de_{\alpha}}{\alpha\in I}$ a collection of bounded
linear operators $\de_{\alpha}\,:\,\A\rar\A$, such that
\item{(1)} $\de_{\alpha}(\idty)=0,\quad \forall\alpha\in I$;
\item{(2)} $\A\fin:=\Set\big{A\in\A}{\quad\sum_{\alpha\in I}
             \Norm\big{\de_{\alpha}(A)}<\infty}$ is $\norm{\cdot}$-dense
      in $\A$;
\item{(3)} There exists a state $\eta\in\A^*$ (reference state), such
      that for $A\in\A\fin$:
$$  \norm{A-\eta(A)\idty}
      \le\sum_{\alpha\in I} \Norm\big{\de_{\alpha}(A)}
\quad. \deqno(osc1)$$

\noindent Then, for $A\in\A$,
$$\trnm{A}:=\sum_{\alpha\in I}\Norm\big{\de_{\alpha}(A)}$$
is called the {\bf oscillation norm} of $A\in\A$.
\eproclaim

Obviously, $\trnm{\cdot}$ is a seminorm on $\A$. It satisfies
$$ \inf_{\lambda\in\Cx}\norm{A-\lambda\idty}\quad
   \leq\quad\norm{A-\eta(A)\idty}           \quad
   \leq\quad\trnm{A}
\quad.\deqno(lowbd)$$
In particular, $\trnm{A}=0$ implies that $A$ is a multiple of the
identity.

A simple argument shows that an infimum such as the one on the left
hand side of \eq(lowbd) is attained at some $\lambda=\lambda(A)$ in
any normed space $\A$ with a fixed element $\idty\in\A$. In a
Hilbert space we can even assert that $\lambda(A)$ is uniquely
determined, and depends linearly on $A$. In a C*-algebra, however,
the shape of the unit ball is different and although, for hermitian
$A$, $\lambda(A)$ is uniquely determined, it is a non-linear
functional. If the overall bound in \eq(lowbd) holds, however, we
can easily make {\it every} state an admissible reference state,
albeit for the oscillation norm $\trnm{A}'=2\trnm{A}$: if $\lambda$
is such that $\norm{A-\lambda\idty}\leq\trnm{A}$, then, for any
state $\eta$,
$\norm{A-\eta(A)\idty}
    =\norm{(A-\lambda\idty)+\eta(A-\lambda\idty)\idty}
    \leq 2\norm{A-\lambda\idty}
    \leq2\trnm{A}$.
This shows that the choice of the reference state is largely
arbitrary. However, the existence of some such state  is a
non-trivial constraint on $\set{\de_\alpha}$, as the following
example shows.

\example/de:qloc/
Let $\A=\bigotimes_{i\in\Ir}\A^i$ be the quasilocal algebra on the
one-dimensional chain, with $\A^i\cong\A^1$ for all $i$. Denote by
$\tau:\A\to\A$ the translation automorphism, defined by
$$  \tau\bigl(\bigotimes_{i\in\Ir} A^i \bigr)
         = \bigotimes_{i\in\Ir} A^{i+1}
\quad.$$
We define an operator $\de:\A\to\A$ by
$$\de(A):=\tau(A) - A
\quad.$$
Then $\de(A)=0$ is equivalent to the translation invariance of $A$,
which in the quasi-local algebra $\A$ is equivalent to $A\in\Cx\idty$.
Thus the one-element collection $\set\de$ satisfies the first two
conditions of \Def/osno/ (with $\A\fin=\A$), and, moreover,
$\de(A)=0$ implies $A\in\Cx\idty$. But condition 3 is
violated in the strong form that there is no finite constant $C$
such that \eq(lowbd) holds as
$\norm{A-\lambda\idty} \leq\,C\trnm{A}$.

To see this, pick some element $A_1\in\A^1\setminus\Cx\idty$ in the
one-site algebra, and a function $f:\Ir\to\Rl$ with $f(i)\geq0$, and
$\sum_if(i)=1$. Then set
$$ A=\sum_if(i)\ \tau^i(A_1)\quad.$$
Let $\omega_1$ be a state on $\A^1$, and let
$\omega_1^\Ir=\bigotimes_{i\in\Ir}\omega_1$ be the
infinite product state on $\A$. Then
$$\eqalign{
   \norm{A-\lambda\idty}
     &\geq \omega_1^\Ir(A-\lambda\idty)
       =\omega_1(A_1)-\lambda  \cr
   \norm{\de(A)}
     &=\Norm\big{\sum_i\bigl(f(i)-f(i-1)\bigr) \tau^i(A_1)}
      \leq \norm{A_1}\ \sum_i\abs{f(i)-f(i-1)}
\quad.}$$
with a suitable choice of $\omega_1$ and $\lambda$ we find
$\norm{A-\lambda\idty}\geq{\inf}_{\lambda'}\norm{A_1-\lambda'\idty}$, and
hence, with $\trnm A=\norm{\de(A)}$
$$ C^{-1}\quad\leq\quad
   {\trnm{A}  \over \norm{A-\lambda\idty}}
   \quad\leq\quad {\norm{A_1} \over
                   \inf_{\lambda'}\norm{A_1-\lambda'\idty} }\
                     \sum_i\abs{f(i)-f(i-1)}
\quad.$$
By choosing $f$ to be slowly varying, \eg
$f(i)=(1-\mu)/(1+\mu)\,\mu^{\abs i}$, for $\mu\to1$, the right hand
side can be made arbitrarily small.
\endexample

Oscillation norms defined with a single $\de_\alpha$ are precisely
those of the form \eq(bogus), \ie $\de(A)=A-\eta(A)\idty$. The lower
bound \eq(osc1) is then equivalent to $\trnm A_0\leq\trnm A$. In the
other direction we have the estimate (see Appendix IV of
\cite{Boguslav})
$$ \trnm A =\sum_\alpha\norm{\de_\alpha(A-\eta(A)\idty)}
        \leq\left(\sum_\alpha\norm{\de_\alpha}\right)
        \trnm A_0
\quad.\deqno(osc1upper)$$
Hence, if there are finitely many oscillation operators
$\de_\alpha$, or the above sum is otherwise convergent, the norms
$\trnm\cdot$ and $\trnm\cdot_0$ are equivalent. This is not a big
surprise, since on a finite dimensional $\A$, all semi-norms which
vanish exactly on the constants are equivalent.

In the next example we will consider a special case of \Def/osno/,
which will sometimes be especially convenient. The oscillation norms
used in \cite{MATA,MATB,MATC} and \cite{Boguslav} are of this form.

\example/matsui/
We say that the operators $\de_\alpha$ satisfy {\it Matsui's
condition}, when
$$\sum_{\alpha\in I}\de_{\alpha}(A)=A - \eta(A)\idty
\quad,\deqno(matsui)$$
for some state $\eta$. Then, by a simple application of the triangle
inequality, condition 3 of \Def/osno/ is satisfied.
Another situation in which this condition comes up naturally is the
following: Let $\A=\M_n$ be the algebra of $n\times n$-matrices, and
let $G$ be a finite group. Consider an irreducible projective
representation $g\mapsto U_g\in\M_n$ of $G$, and set
$$  \de_g(A)={1\over\abs G}\bigl(A-U_gAU_g^*\bigr)
\quad.\deqno(deGroup)$$
Then $\abs G^{-1}\sum_gU_gAU_g^*$ commutes with all $U_g$, and is
hence of the form $\eta(A)\idty$. Clearly, $\eta$ is an invariant
state with respect to all $U_g$, and must therefore be the
normalized trace of $\M_n$. Since for the identity element $e\in G$
we have $\de_e=0$, it suffices to take the above $\de_g$ for
$g\in I=G\setminus\set e$. With this choice, Matsui's condition
\eq(matsui) holds. Note that since $\norm{A}=\norm{AU_g}$ the
oscillation norm may be written in the suggestive form
$$ \trnm A={1\over\abs G}\sum_{g\in G} \Norm\big{\bracks{A,U_g}}
\quad.$$
The simplest special case is to take the $U_g$ as
the three Pauli matrices $\sigma_\alpha\in\M_2$, and $U_e=\idty$:
the product of any two of these operators is in the same set, up to
a phase. The group $G$ ($\abs G=4)$ consists of the rotations by
$\pi$ around the three Cartesian axes in $\Rl^3$. In this case one
also finds easily that the reference state $\eta$ is uniquely
determined: let $\eta'$ be another reference state. Then
$$ \norm{\sigma_1-\eta'(\sigma_1)\idty}
     \leq\trnm{\sigma_1}
     ={1\over4}\sum_{i=1}^3\Norm\big{\bracks{\sigma_1,\sigma_i}}
     ={1\over4}(\norm{2\sigma_3}+\norm{2\sigma_2})
     =1
$$
implies that $\eta'(\sigma_1)=0$. Repeating the same argument for the
other components we find that $\eta'=\eta$ has to be the normalized
trace.
\endexample

The infimum on the left hand side of \eq(lowbd) is the standard
quotient norm of $\A/\Cx\idty$. Hence $\trnm\cdot$ can be considered
as a proper norm on the subspace $\A\fin/\Cx\idty$ of this quotient.
We now show that this norm turns $\A\fin/\Cx\idty$ into a Banach
space.

\iproclaim//Lemma.
$\A\fin/{\Cx\idty}$ is $\trnm{\cdot}$-complete.
\eproclaim

\proof:
Let $\set{A_n}\subset\A\fin/\Cx\idty$ be a $\trnm{\cdot}$-Cauchy
sequence. It follows that $\norm{A_n}\le\trnm{A_n}$. Because
$\A/\Cx\idty$ is complete for the quotient norm  $\norm{\cdot}$,
there exists an $A\in\A/\Cx\idty$ such that $\norm{A-A_n}\to0$ for
$n\to\infty$. We have to show that $A\in\A\fin/\Cx\idty$. Let
$I'\subset I$ be a finite subset of the index set. Then, because
each $\de_\alpha$ is bounded,
$$\eqalign{
  \sum_{\alpha\in I'} \Norm\big{\de_{\alpha}(A)}
  &= \lim_{n\to\infty} \sum_{\alpha\in I'}
      \Norm\big{\de_{\alpha}(A_n)} \cr
  &\le \limsup_{n\to\infty} \sum_{\alpha\in I}
      \Norm\big{\de_{\alpha}(A_n)} \cr
  &\le \lim_{n\to\infty} \trnm{A_n} .
}$$
This limit exists because $A_n$ is $\trnm\cdot$-Cauchy. Taking the
supremum over all finite $I'\subset I$, we find that
$\trnm{A}<\infty$. Similarly, we find that
$$ \sum_{\alpha\in I'} \Norm\big{\de_{\alpha}(A-A_n)}
    \leq \lim_{m\to\infty}\trnm{A_m-A_n}
\quad,$$
and since this bound is independent of $I'\subset I$, we have that
$\lim_n\trnm{A-A_n}=0$.
\QED

Since we have not postulated any further properties of the operators
$\de_\alpha$, the space $\A\fin$ does not come with a natural
algebraic structure. However, in the special case, where
$\de_\alpha(A)=i\bracks{A,D_\alpha}$ are derivations, but also in
the case \eq(deGroup), we get
\break
$\norm{\de_\alpha(AB)}
   \leq \norm{\de_\alpha(A)}\norm{B}
       +\norm{A}\norm{\de_\alpha(B)}$,
and hence
$$ \trnm{AB}\leq \trnm A\norm B+ \norm{A}\trnm B
\quad. \deqno(deriv)
$$
In particular, $\A\fin$ becomes a Banach algebra with the norm
$\norm{A}_\lambda=\norm{A}+\lambda\trnm A$, for any $\lambda>0$.

The main reason for introducing oscillation norms is that in the
case of large systems it is often easier to establish contractivity
in this norm than in the C*-norm $\norm{\cdot}$. Nevertheless, as
the following Proposition shows, this contractivity is sufficient to
establish convergence of the iterates in the C*-norm, and hence
ergodicity. Note that while the oscillation norms \eq(bogus),
\eq(strnm), \eq(Krist), and \eq(sumtrnm) are ``equivalent'' in finite
dimensional situations, the best constant $\ep$ in the assumption of
the Proposition will depend on the choice of norm.

\iproclaim/ergodicity/Proposition.
Let $\A$ be a C*-algebra with oscillation norm $\trnm\cdot$, and
consider a linear operator $P:\A\rar\A$ such that
$\norm{P(A)}\le\norm{A}$,
$P(\idty)=\idty$, and, for some fixed $\ep<1$, and all $A\in\A\fin$,
$$\trnm{P(A)}\le\ep\,\trnm{A}
\quad.$$
Then there exists a unique state $\rho\in\A^*$, such that
$\rho\circ P=\rho$. Moreover,
\item{(1)} $\lim_{n\to\infty}\norm{P^n(A) - \rho(A)\idty}=0$, for all
      $A\in\A$, and
\item{(2)} $\norm{P^n(A) - \rho(A)\idty}\le 2\,\ep^n \trnm{A}$,
      for all $A\in\A\fin$.
\eproclaim

\proof:
Fix an element $A\in\A\fin$ and consider the sequence
$A_n=P^n(A)$. Choose some numbers $\lambda_n$, for example
$\lambda_n=\eta(A_n)$, such that
$$\Norm\big{A_n - \lambda_n\idty}
  \le\trnm{A_n}
\quad. $$
Then $\Set\big{:}{\lambda_n}$ is a Cauchy sequence: taking $n\geq m$
without loss of generality, we get
$$\eqalign{
  \abs{\lambda_n - \lambda_m} &\le
  \norm{\lambda_n\idty - A_n} + \norm{A_n - \lambda_m\idty} \cr
  &\le \trnm{A_n} + \norm{P^{n-m}}\,\norm{A_m - \lambda_m\idty}\cr
  &\le \bigl(\ep^n + \ep^m \bigr)\,\trnm{A}
\quad.}$$
Let $\tilde{\rho}(A)=\lim_{m\to\infty}\lambda_m$. Inserting this
definition into the previous inequality, we find
$$\abs{\lambda_n - \tilde{\rho}(A)}\le\ep^n \,\trnm{A}
\quad,$$
and for $A\in\A\fin$ it follows that
$$\eqalign{
  \norm{P^n(A) - \tilde{\rho}(A)\idty} &\le
  \norm{P^n(A) - \lambda_n\idty} + \abs{\lambda_n - \tilde{\rho(A)}} \cr
  &\le \trnm{A_n} + \ep^n\trnm{A} \cr
  &\le 2 \ep^n \trnm{A}
\quad.}$$
$\tilde{\rho}(A)$ is linear, because $A\mapsto \tilde\rho(A)\idty$ is
the limit of the linear operators $P^n$.
$\tilde{\rho}(A)$ is also a $\norm{\cdot}$-continuous functional on
$\A\fin$:
$$\eqalign{
    \abs{\tilde{\rho}(A)} &\le
    \norm{\tilde{\rho}(A)\idty - P^n(A)} + \norm{P^n(A)}\cr
    &\le 2\,\ep^n\,\trnm{A} + \norm{A}
    {\stack{n\to\infty}{\longrightarrow}} \norm{A}.
}$$
Let $\rho$ denote the continuous extension of $\tilde{\rho}$ on
$\A$. Then $\abs{\rho(A)}\leq\norm{A}$, \ie $\norm{\rho}=1$, and
also $\rho(\idty)\idty=\lim_nP^n(\idty)=\idty$. Hence $\rho$ is a
state.

For showing the convergence (1), let $A\in\A$, and $\delta>0$. Then
we may pick $A_\delta\in\A\fin$ with
$\norm{A-A_\delta}\leq\delta$. It follows that
$$\eqalign{
 \norm{P^n(A) - \rho(A)\idty}
   &\le \norm{P^n(A-A_\delta)} + \norm{P^n(A_\delta)
       -\rho(A_\delta)\idty} +     \abs{\rho(A_\delta-A)}\cr
   &\le \norm{A-A_\delta} + 2\,\ep^n\trnm{A_\delta}
       + \norm{A_\delta-A} \cr
   &\le 2\delta+ 2\,\ep^n\trnm{A_\delta}
\quad.}$$
Hence, for sufficiently large $n$, the left hand side becomes
arbitrarily small.

Finally, suppose $\nu\in\A^*$ is another fixed point of the adjoint
of $P$, \ie $\nu\circ P=\nu$. Then
$$ \nu(A)=\lim_{n\to\infty} \bigl(\nu\circ P^n\bigr) (A)
         =\nu\bigl(\rho(A)\idty\bigr)
         =\rho(A)\cdot \nu(\idty)
\quad.$$
Hence, if $\nu$ is also normalized, we have $\nu=\rho$.
\QED

We want to use this criterion to establish ergodicity of cellular
automata describing spin systems. The operator $P$ will then be built
up from local operators $P_x$. We thus have to construct an
oscillation norm for composite systems, \ie for tensor products
of algebras with oscillation norm, and then have to apply
\Prp/ergodicity/ to the product system.

\bgssection 3.2. Tensorable oscillation norms

We now take the algebra $\A=\A^\La$ to be the quasi-local algebra
over a lattice $\La$, to each site of which is attached a unital
C*-algebra $\A^x$, \ie
$$ \A\equiv\A^\La
      =\bigotimes_{x\in\La} \A^x
\quad.$$
Assume now that we are given an oscillation norm for the algebras
$\A^x$ at each site. We would like to assemble from this an
oscillation norm for $\A^\La$.

The basic idea is very simple: as the collection of ``flip''
operators we simply take the union of the flip operators for each
site, \ie
$$\bigcup_{x\in\La}\Set\big{\de_{\alpha}^{(x)}}{\alpha\in I^x}
\quad,\deqno(osctensor1)$$
where
$\de_{\alpha}^{(x)}
    = \de_{\alpha}^{x}\otimes\id^{\La\bs\set{x}}$
is just the action of $\de_{\alpha}^x$ at site $x$ of the quasilocal
algebra. The oscillation norm of an element $A\in\A$ is then
$$ \trnm{A} = \sum_{x\in\La} \sum_{\alpha\in I^x}
  \Norm\big{\de_{\alpha}^{(x)}(A)}
\quad.\deqno(tensorTrnm)$$
This quantity is always defined, but possibly infinite. So we can
define, as before,
$$ \A\fin
   = \Set\big{A\in\A}{\trnm A<\infty}
\quad.$$
Clearly, the finite tensor products $\bigotimes_{x\in\Lambda}A_x$,
with $A_x\in\A\fin^x$, and their linear combinations are in
$\A\fin$. By \Def/osno/, $\A\fin^x\subset\A^x$ is norm dense, and by
definition of the quasi-local algebra the finite tensor products
span a dense subspace of $\A$.  Hence $\A\fin\cap\A\loc$, and a
fortiori $\A\fin$, is dense in $\A$. In order for $\trnm\cdot$ to
become an oscillation norm, we need to establish the estimate
\eq(osc1) in the definition of oscillation norms, with the obvious
candidate
$$\eta= \bigotimes_{x\in\La}\eta^x
\deqno(tensoref)$$
for a reference state. For this, we need the following ``stabilized
version'' of \eq(osc1). It is automatically satisfied in
the commutative case (cf.\ \Prp10/cl=tensorable/ below). For the
notion of complete boundedness, see the Appendix.  When $\eta$ is a
state on the C*-algebra $\A$, we denote by $\eth:\A\to\A$ the
operator $\eth(A)=\eta(A)\idty$.

\vtop{
\iproclaim/osctensor2/Definition.
An oscillation norm defined by operators
$\Set\big{\de_{\alpha}}{\alpha\in I}$ on a C*-algebra $\A$, with
reference state $\eta$, is called  {\bf tensorable}, if each
$\de_{\alpha}$ is completely bounded, and
$$\Norm\big{A - \bigl(\id_{\M_n}\otimes\eth\bigr)(A)}
  \le\sum_{\alpha\in I}\Norm\big{\bigl(\id_{\M_n}
  \otimes\de_{\alpha}\bigr)(A)}
\quad,\deqno(oscten)$$
for $A\in\M_n(\Cx)\otimes\A$.
\eproclaim
}

At first sight it may seem rather special to allow tensoring only
with $\M_n$. However, as in the definition of complete positivity
this is sufficient to give the analogous statement for all
C*-algebras.

\iproclaim/allM/Lemma.
Let  $\Set\big{\de_{\alpha}}{\alpha\in I}$ define a tensorable
oscillation norm on a C*-algebra $\A$, and let $\M$ be any unital
C*-algebra. Let $\M\otimes\A$ be  the minimal C*-tensor product, and
$A\in\M\otimes\A$. Then
$$\Norm\big{A - \bigl(\id_{\M}\otimes\eth\bigr)(A)}
  \le\sum_{\alpha\in I}\Norm\big{\bigl(\id_{\M}
  \otimes\de_{\alpha}\bigr)(A)}
\quad.\deqno\lasteq'(osctenA)$$
\eproclaim

\proof:
\def\phg{\widehat p_\gamma}%
We need the following basic observation: if $p_\gamma$ is a
net of Hilbert space operators with $\norm{p_\gamma}\leq1$,
converging strongly to the identity operator then, for any bounded
operator $A$,
$$ \lim_\gamma\norm{p_\gamma^* Ap_\gamma}=\norm{A}
\quad.\deqno(comprnorm)$$
Indeed, the inequality
$\limsup_\gamma\norm{p_\gamma^* Ap_\gamma}
    \leq\limsup_\gamma\norm{p_\gamma}^2\norm{A}
    \leq\norm{A}$
is trivial. On the other hand, let $\phi,\psi$ be unit vectors such
that $\abs{\bra\phi,A\psi>}\geq\norm{A}-\epsilon$, and let $\gamma$ be
sufficiently large such that $\norm{p_\gamma\phi-\phi},\
\norm{p_\gamma\psi-\psi}\leq\epsilon$. Then
$$\norm{p_\gamma^* Ap_\gamma}
      \ \geq\ \abs{\bra p_\gamma\phi,Ap_\gamma\psi>}
      \ \geq\ \abs{\bra\phi,A\psi>} -2\epsilon\norm{A}
      \ \geq\  \norm{A}-\epsilon-2\epsilon\norm{A}
\quad,$$
and \eq(comprnorm) follows by taking the inferior limit, and
$\epsilon\to0$.

Without loss we may take $\M$ and $\A$ to be faithfully represented
on some Hilbert spaces. Then the minimal C*-tensor product
$\M\otimes\A$ is defined as the C*-algebra generated by operators of
the form $M\otimes A$, with $M\in\M$ and $A\in\A$, respectively. Let
$p_\gamma$ denote a net of finite dimensional projections in the
representation space of $\M$, converging to the identity, and
introduce the operators $\phg(M)=p_\gamma Mp_\gamma$. Then, for
$A\in\M\otimes\A$, we have
$\widehat A=(\phg\otimes\id_\A)(A)\in\M_n\otimes\A$,
where $n$ is the dimension of $p_\gamma$. Hence, because
$(\phg\otimes\id_\A)$ and $(\id_{\M}\otimes\eth\bigr)$ commute,
$$\eqalign{
    \Norm\big{(\phg\otimes\id_\A)
             (A - \bigl(\id_{\M}\otimes\eth\bigr)(A))}
    &=\Norm\big{\widehat A - \bigl(\id_{\M_n}\otimes\eth\bigr)
               (\widehat A)} \cr
    &\leq \sum_{\alpha\in I}\Norm\big{
                    \bigl(\id_{\M_n}\otimes\de_{\alpha}\bigr)
                         (\widehat A)}\cr
    &= \sum_{\alpha\in I}\Norm\big{
               \bigl(\phg\otimes\id_\A\bigr)
               \bigl(\id_{\M}\otimes\de_{\alpha}\bigr)(A)}\cr
    &\leq\sum_{\alpha\in I}\Norm\big{
               \bigl(\id_{\M}\otimes\de_{\alpha}\bigr)(A)}
\quad.\cr}$$
Hence the result follows, by applying \eq(comprnorm) to the left
hand side of this inequality, and the family of projections
$p_\gamma\otimes\idty$.
\QED

The basic result concerning the tensor product of algebras with
oscillation norm is the following. The second part was proven in a
special case by Matsui \cite{MATD}.

\iproclaim/TensorableOscillation/Proposition.
Let $\Set\big{\A^x}{x\in\La}$ be algebras with ten\-sor\-able
oscillation norms, defined by $\Set\big{\de_{\alpha}^x}{\alpha\in
I^x}$, and with reference states $\eta^x\in\bigl(\A^x\bigr)^*$. Then
\item{(1)}
$\A=\bigotimes_{x\in\La}\A^x$
is an algebra with tensorable oscillation norm given by
\eq(osctensor1), with reference state \eq(tensoref).
\item{(2)}
$\A\fin\cap\A\loc$ is dense in $\A\fin$ with respect to
$\trnm\cdot$.
\eproclaim

\proof:
We show (2) first.
For any finite subset $\Lambda\subset\La$, define
$$\eth^{\Lambda^c}:=\prod_{x\notin\Lambda} \eth^{(x)}
\quad,$$
where
$\eth^{(x)}(A)=\bigl(\eth^x\otimes\id^{\La\bs\set{x}}\bigr)(A)$.
Therefore the range of $\eth^{\Lambda^c}$ is contained in the local
algebra $\bigotimes_{x\in\Lambda}\A^x$.
For $A\in\A\loc$, we trivially have
$$\Norm\big{\eth^{\Lambda^c}(A) - A}
  \stack{\Lambda\to\La}{\longrightarrow} 0
\quad,\deqno*()$$
because $\Lambda$ absorbs the localization region of $A$.
Consequently, because $\A\loc\subset\A$ is $\norm{\cdot}$-dense,
\eq*() is valid for all $A\in\A$.

We show next that
$$\trnm{\eth^{\Lambda^c}(A) - A}
  \stack{\Lambda\to\La}{\longrightarrow} 0
\quad,$$
for $A\in\A\fin$. One has
$$\eqalign{
  \trnm{\eth^{\Lambda^c}(A) - A}
   &= \sum_{x\in\Lambda} \sum_{\alpha\in I^x}
        \Norm\big{\de_{\alpha}^{(x)} \bigl(\eth^{\Lambda^c}-\id^{\La}
        \bigr) (A)} \cr
  &= \sum_{x\in\Lambda} \sum_{\alpha\in I^x}
        \Norm\big{\de_{\alpha}^{(x)} \bigl(\eth^{\Lambda^c}-\id^{\La}
        \bigr) (A)}
    + \sum_{x\notin\Lambda} \sum_{\alpha\in I^x}
        \Norm\big{\de_{\alpha}^{(x)} \bigl(\eth^{\Lambda^c}-\id^{\La}
        \bigr) (A)} \cr
  &= \sum_{x\in\Lambda} \sum_{\alpha\in I^x}
        \Norm\big{\bigl(\eth^{\Lambda^c}-\id^{\La} \bigr)
                      \de_{\alpha}^{(x)}(A)}
    + \sum_{x\notin\Lambda} \sum_{\alpha\in I^x}
        \Norm\big{\de_{\alpha}^{(x)} (A)}
}$$
because for $x\in\Lambda$ the operators $\eth^{\Lambda^c}$ and
$\de_{\alpha}^{(x)}$ commute, and for $x\notin\Lambda$,
$\de_{\alpha}^{(x)}$ acts on the factor $\idty^{(x)}$, so
$\de_{\alpha}^{(x)}\circ\eth^{\Lambda^c}=0$.
In the limit $\Lambda\to\La$ the
second term vanishes because the sum over all $x\in\La$ converges
for $A\in\A\fin$. The first sum is termwise dominated by the sum
defining $2 \trnm{A}$. However, because
$\de_{\alpha}^{(x)}(A)\in\A$, each term in this sum goes to zero by
\eq*(). Hence the sum goes to zero by dominated convergence.
Hence $\trnm{\eth^{\Lambda^c}(A) - A}\to0$, and $\A\fin\cap\A\loc$
is $\trnm\cdot$-dense in $\A\fin$.

To prove (1), we have to verify \Def/osno/. The boundedness of the
$\de_{\alpha}^{(x)}$ follows from the complete boundedness of the
operators $\de_{\alpha}^{x}$ on $\A^x$. The property
$\de_{\alpha}^{x}(\idty)=0$ is evident, and we argued for the
$\norm{\cdot}$-density of $\A\fin$ above, before \eq(tensoref).
Hence only the estimate \eq(osc1) remains to be seen.
We begin by showing it for $A\in\A\fin\cap\A\loc$, say $A$ localized
in a finite region $\Lambda\subset\La$. We conveniently label the
sites in $\Lambda$ as $1,\ldots,\abs\Lambda$. Then with a
telescoping sum we find
$$\eqalign{
 \Norm\Big{A-\bigotimes_{x\in\Lambda}\eth^x(A)}
   &\le  \sum_{k=1}^{\abs{\Lambda}}
       \Norm\Big{\bigotimes_{x<k}\eth^x \otimes
            \bigl(\id^k - \eth^k\bigr) \otimes
            \bigotimes_{y>k}\id^y \,(A)} \cr
    &= \sum_{k=1}^{\abs\Lambda}
        \Norm\Big{\prod_{x<k}\eth^{(x)}}  \,
        \Norm\big{A -\eth^{(k)}(A)}      \cr
    &\leq \sum_{k\in\Lambda} \sum_{\alpha\in I^k}
            \Norm\big{\de_{\alpha}^{(k)}(A)}
     = \trnm{A}
\quad.}$$
where $\eth^{(x)}:\A\to\A$ is the operator
$\eth^x\otimes\id^{\La\setminus\set x}$.  At the last estimate we
used that the product of the $\eth^{(x)}$ is a completely positive
unit preserving operator, which has hence norm $1$, and, of course,
the tensorability of $\trnm\cdot$. Hence the required estimate holds
for $A\in\A\fin\cap\A\loc$. In particular, it holds for the
approximants $\eth^{\Lambda^c}(A)$ of a general $A\in\A\fin$. Since
by part (2) of the Proposition these elements approximate $A$ in both
norms, both sides of the estimate converge as $\Lambda\to\La$. This
completes the proof that $\trnm\cdot$ is an oscillation norm. It is
tensorable, because we can include an additional tensor factor
$\M_n$ in the product defining $\A$, and use the same estimates as
above to establish \eq(oscten).
\QED

With this Proposition the construction of oscillation norms for
infinite systems is reduced to the construction of tensorable
oscillation norms for the one-site algebras. Recall from
\eq(osc1upper) that all oscillation norms for which
$\sum_{\alpha\in I}\norm{\de_\alpha}<\infty$ are equivalent.
By summing this estimate over all cells we conclude similarly that
all oscillation norms on a composite system, for which
$$ \sum_{\alpha\in I^x}\cbnorm{\de_\alpha^x}\leq c<\infty
\deqno()$$
with a constant independent of $x$, are also equivalent. In
particular, this holds for the norm in \cite{Boguslav}. Equivalence
can also be shown for global oscillation norms, which are defined as
the sum of local oscillations, which are in turn given by some
supremum (see \eq(strnm) or \eq(Krist)) \cite{Krist}.

Showing tensorability is especially easy in the classical case:

\iproclaim/cl=tensorable/Proposition.
On an abelian C*-algebra $\A$, every oscillation norm is tensorable.
\eproclaim

\proof:
We can set $\A=\C(X)$, for some compact space $X$.
Let $\M$ be any C*-algebra. Then
$\M\otimes\A\cong \C(X,\M)$, the algebra of $\M$--valued continuous
functions on $X$, with the norm
$$ \norm{A}:=\sup_x\norm{A(x)}=\sup_{x,\Phi}\abs{\Phi(A(x))}
\quad,\deqno(normCXM)$$
where the supremum over $\Phi$ is with respect to all linear
functionals on $\M$ with $\norm{\Phi}\leq1$.
Then
$\bigl((\id_\M\otimes\de_\alpha)A\bigr)(x)
    =\int\de_\alpha(x,dy)\ A(y) $,
where we have used $\de_\alpha$ to denote both the operator and its
integral kernel. Applying the estimate \eq(osc1) to the continuous
function $x\mapsto\Phi(A(x))$, we obtain
$$\eqalign{
  \norm{A-(\id_\M\otimes\eth)(A)}
       &=\sup_{x,\Phi}\abs{\Phi(A(x))-\int\eta(dy)\Phi(A(y))} \cr
       &\leq\sup_\Phi\, \sum_{\alpha\in I} \sup_x
          \abs{\int\de_\alpha(x,dy)\ \Phi(A(y))}  \cr
       &\leq\sum_{\alpha\in I}\sup_\Phi\,  \sup_x
          \abs{\int\de_\alpha(x,dy)\ \Phi(A(y))}  \cr
       &=\sum_{\alpha\in I} \norm{(\id_\M\otimes\de_\alpha)(A)}
\quad.}$$
\QED

In the non-commutative case, tensorability is a non-trivial
constraint on oscillation norms. The following is a handy criterion.

\iproclaim/tensorMatsui/Lemma.
Suppose that the operators $\Set\big{\de_{\alpha}}{\alpha\in I}$
defining an oscillation norm $\trnm\cdot$ on a C*-algebra $\A$
satisfy Matsui's condition (see \Exa/matsui/ above).
Then it is tensorable.
\eproclaim

\proof:
Let $A\in\M_n(\Cx)\otimes\A$. Then, by \eq(matsui),
$$\eqalign{
  \sum_{\alpha\in I}\bigl(\id_{\M_n}\otimes\de_{\alpha}\bigr)(A)
  &= \Bigl( \id_{\M_n}\otimes\sum_{\alpha\in I}\de_{\alpha}
      \Bigr) (A)
  = \Bigl( \id_{\M_n}\otimes \bigl(\id_{\A} - \eth\bigr)
      \Bigr) (A)\cr
  &= A - \Bigl( \id_{\M_n}\otimes \eth\Bigr) (A)
\quad.}$$
By taking norms on both sides we find \eq(oscten).
\QED

Finally, we record for later use that for tensor product operators
a version of \eq(deriv) holds without further assumptions: using
that in a minimal C*-algebra tensor product
$\norm{A\otimes B}=\norm{A}\,\norm{B}$, we find
$$ \trnm{A\otimes B}=\trnm{A}\,\norm{B}+\norm{A}\,\trnm{B}
\quad.\deqno(LocalOsc)$$

\bgsections 4.   Contractivity of tensor product operators
\bgssection 4.1. Oscillation norm estimate

In the simplest, non-interacting case of a cellular automaton the
total transition operator $P$ is the infinite tensor product of the
one-site transition operators $P_x$. If we know that each $P_x$
contracts exponentially with rate $\epsilon<1$ to a multiple of the
identity, can we also assert this about $P$? This turns out to be
the crucial question for developing ergodicity estimates for quantum
cellular automata.
We will show in this subsection that, provided we use oscillation
norms to express contractivity, the product $P$ indeed contracts
with the same rate. In fact the validity of this bound is the main
reason for considering oscillation norms. In order to make this
point more precise we show in the next subsection how other norms
fail to give the desired estimate.

The most straightforward definition of contractivity of a transition
operator $P$ is given by the estimate $\trnm{P(A)}\le\ep\,\trnm{A}$
as in \Prp/ergodicity/. The best constant $\ep$ in this estimate is
conveniently denoted by $\trnm P$; it is the norm of $P$ as an
operator on $\A\fin$. However, this quantity is not appropriate for
the study of composite systems, since even in the classical case we
may have $\trnm{\id_\M\otimes P}>\trnm{P}$ (cf.\ the example at the
end of this subsection). Therefore, we use a version of $\trnm P$,
which is ``stabilized'' with respect to coupling the system to an
outside world.

\iproclaim/cbOscillationNorm/Definition.
Let $\A$ and $\B$ be algebras with tensorable oscillation norms
ge\-ne\-ra\-ted by $\Set\big{\de_{\alpha}}{\alpha\in I}$ and
$\Set\big{\tilde{\de}_{\beta}}{\beta\in{J}}$,
respectively. Let $P:\A\rar\B$ be linear and completely bounded.
Then the {\bf completely bounded oscillation norm} of $P$, denoted
by $\trnmcb{P}$, is defined as the smallest constant $\ep$ for which
the inequality
$$  \sum_{\beta\in J}
  \Norm\Big{\bigl(\id_{\M_n}\otimes(\tilde{\de}_{\beta}P)\bigr)(A)}
  \leq \ep\, \sum_{\alpha\in I}\Norm\big{\bigl(\id_{\M_n}
                            \otimes\de_{\alpha}\bigr)(A)}
\deqno(trnmcb)$$
holds, for all $n\in\Nl$, and $A\in\M_n\otimes\A$.
\eproclaim

In particular, for $n=1$, we get $\trnm{PA}\leq\trnmcb{P}\,
\trnm A$, \ie $\trnm P\leq\trnmcb P$. Precisely as in \Lem/allM/ one
sees that if the bound of the form \eq(trnmcb) holds for all
algebras $\M_n\otimes\A$, it also holds for all minimal C*-tensor
products $\M\otimes\A$ with other C*-algebras. The crucial property
of this norm is given in the following Theorem.

\iproclaim/contractensor/Theorem.
Let $P_x:\A^x\to\B^x$, $x\in\La$ be a family of completely positive
unit preserving operators between algebras with tensorable
oscillation norms. Let $\A=\bigotimes_{x\in\La}\A^x$ and
$\B=\bigotimes_{x\in\La}\B^x$ be equipped with the oscillation norm
described by \Prp/TensorableOscillation/, and let $P:\A\to\B$ be
defined by  $P=\bigotimes_{x\in\La}P_x$. Then
$$ \trnmcb{P}=\sup_{x\in\La}\trnmcb{P_x}
\quad.\deqno(contractensor)$$
\eproclaim

\proof:
The inequality $\trnmcb P\geq\sup_x\trnmcb{P_x}$ is trivial, because
the estimate \eq(trnmcb), written for $P$, and an observable
$A\in\idty_\M\otimes\A^x\subset\idty_\M\otimes \A$ reduces to the
corresponding estimate for $P_x$.

For the opposite inequality we have to show that, provided each
$P_x$ satisfies the estimate \eq(trnmcb) with the same constant
$\ep$, \ie $\ep\geq\trnmcb{P_x}$ for all $x\in\La$, then so does
$P$. Consider $A\in\M\otimes\A$, and one term in the sum on the left
hand side of \eq(trnmcb), say for $x\in\La$, and $\beta\in J^x$. We
have to estimate
$$ \bigl(\id_\M\otimes (\tilde\de^{(x)}_\beta P)\bigr)(A)
     =\Bigl(\id_\M\otimes \bigotimes_{y\neq x}P_y\
        \otimes\id_{\A^x}\Bigr)
      \Bigl(\id_\M\otimes \bigotimes_{y\neq x}\id_{\A^y}
                  \otimes (\tilde\de_\beta^x P_x)
            \Bigr)(A)
\quad.$$
The first parenthesis is a contraction because each $P_y$ is
completely positive and unital. Hence taking norms, summing over
$\beta\in J^x$, and using \Def/cbOscillationNorm/, we find
$$\eqalign{
  \sum_{\beta\in J^x}\Norm\Big{
        \bigl(\id_\M\otimes (\tilde\de^{(x)}_\beta P)\bigr)(A)}
   &\leq\trnm{P_x}\sum_{\alpha\in I^x}
         \Norm\Big{(\id_\M\otimes \bigotimes_{y\neq x}\id_{\A^y}
                  \otimes \de_\alpha^x)(A)}       \cr
   &\leq\ep\sum_{\alpha\in I^x}
      \Norm\Big{\bigl(\id_\M\otimes\de^{(x)}_\alpha\bigr)(A)}
\quad.}$$
The sum of these inequalities over all lattice points $x\in\La$ is
the desired estimate, showing $\trnmcb{P}\leq\ep$.
\QED

Hence, for a non-interacting system, it suffices to show
$\trnmcb{P_x}\leq\ep<1$ to conclude ergodicity from
\Prp/ergodicity/. However, the explicit estimate of
$\trnmcb{P_x}$ may still be a difficult problem. One way to
handle it is the following {\it decomposition property}.

\vtop{\noindent
\iproclaim/decomposition/Lemma.
In the setting of \Def/cbOscillationNorm/, suppose that for each
$\beta\in J$, we have a decomposition
$$ \tilde\de_\beta\circ P = \sum_{\alpha\in I}
   G_{\beta\alpha} \circ\de_{\alpha}
\quad,$$
with $G_{\beta\alpha}:\A\to\B$ linear and completely bounded, and
the sum strongly convergent on $\A$.  Then
$$ \trnmcb{P}
       \leq \sup_{\alpha\in I}  \sum_{\beta\in J}
    \Norm\big{G_{\beta\alpha}}_{\rm cb}
\quad, \deqno(ContractionRate)$$
where $\cbnorm{\cdot}$ is the completely bounded norm.
\eproclaim
}

\proof:
The proof is obvious by inserting the decompositions of
$\tilde\de_\beta P$, and using the triangle inequality.
\QED

If $\A$ and $\B$ are finite dimensional, such  decomposing maps
$G_{\beta\alpha}$ always exist. In fact, the necessary and
sufficient condition for an operator (here $\tilde\de_\beta P$) to allow
a decomposition with given $\de_\alpha$ is that
$\ker \tilde\de_\beta P\supset\bigcap_\alpha \ker\de_\alpha$. However, the
right hand side of this inclusion is equal to $\Cx\idty$ by
part 1 of \Def/osno/, and, clearly,
$\Cx\idty\subset\ker\tilde\de_\beta P$. We conclude that, on finite
dimensional algebras with oscillation norms defined by finitely
many $\de_\alpha$, we have $\trnmcb P<\infty$ for all transition
operators.

If $\trnmcb P=0$, we must have $\tilde\de_\beta(P(A))=0$ for all
$\beta$, so that $P(A)\equiv P_\omega(A)=\omega(A)\idty$ for some
state $\omega$ on $\A$. Near maps of this form we find transition
operators with small oscillation norms: these could be called
transition operators with {\it large disorder}, since in one step
they wipe out nearly all memory of previous states. It is
straightforward to see from the definition of the completely bounded
oscillation norm that, for $0\leq\lambda\leq1$, the  equation
$$ \trnmcb{(1-\lambda)P_\omega+\lambda P}
       =\lambda\trnmcb P
\deqno(disorder)$$
holds. (In fact, the inequality``$\leq$'' already follows from the
convexity of $\trnmcb\cdot$). Hence, for sufficiently small
$\lambda$, the ergodicity criterion \Prp/ergodicity/ applies, and
$(1-\lambda)P_\omega+\lambda P$ has a unique invariant state, which
will be close to, but not equal to $\omega$.

In some cases, one can use the freedom of adapting the operators
$\de_\alpha$ to the problem at hand to give a simple estimate of
$\trnmcb P$. An example is the following:

\iproclaim/spectralP/Lemma.
Let $P:\A\to\A$ be a transition operator on a finite dimensional
C*-algebra, and suppose that $P$ is diagonalizable, \ie it has a
representation in the form
$$P=D_0 + \sum_{\alpha=1}^N\lambda_\alpha D_\alpha
\quad,$$
with $D_\alpha D_\beta=\delta_{\alpha\beta}D_\alpha$, for
$0\leq\alpha,\beta\leq N$, and $D_0(A)=\omega(A)\idty$ for some
state $\omega$ on $\A$. Define an oscillation norm by setting
$\de_\alpha=D_\alpha$ for $\alpha=1,\ldots,N$. Then
$$ \trnmcb P=\max\Set\Big{\abs{\lambda_\alpha}}{\alpha=1,\ldots,N}
\quad.$$
\eproclaim

\proof:
Note that the oscillation norm so defined satisfies Matsui's
condition \eq(matsui), because $\sum_{\alpha=0}^ND_\alpha=\id$, and
is hence tensorable. The lower bound $\trnmcb
P\geq\max\abs{\lambda_\alpha}$ follows by inserting the eigenvectors
into the estimate defining $\trnmcb\cdot$, and the upper bound
follows from \Lem/decomposition/ with
$G_{\beta\alpha}=\lambda_\alpha \delta_{\alpha\beta}\id $.
\QED

In the classical case we may simplify the definition of
$\trnmcb\cdot$, by considering couplings to classical systems only.
For Ising systems, \ie the case considered in \cite{Lebowitz} and
other principal papers on the subject, the stabilization can even be
omitted entirely.

\iproclaim/classcb/Lemma. Let $\A,\B$ be abelian algebras with
oscillation norm, and $P:\A\to\B$ a completely bounded linear map.
Then
\item{(1)}
$\trnmcb{P}$  is the best constant $\ep$ such that the
estimate
$$  \sum_{\beta\in J}
  \Norm\Big{\bigl(\id_{\M}\otimes(\tilde{\de}_{\beta}P)\bigr)(A)}
  \leq \ep\, \sum_{\alpha\in I}\Norm\big{\bigl(\id_{\M}
                            \otimes\de_{\alpha}\bigr)(A)}
$$
holds for all {\bf abelian} C*-algebras $\M$.
\item{(2)}
if $\B$ is two-dimensional, or the oscillation norm on $\B$ is
defined by a single $\de$, \ie $J$ has only one element, then
$\trnmcb{P}=\trnm{P}$, \ie the best constant is already achieved by
taking $\M$ one-dimensional.
\eproclaim

\proof:
As in the proof of  \Prp/cl=tensorable/ the quantum observable
algebra is reduced to a classical one by evaluating in appropriate
states. Let $\A=\C(X)$, $\B=\C(Y)$, and let $\M$ denote the
C*-algebra of bounded functions on the index set $J$. Suppose the
bound in (1) holds with this abelian algebra $\M$, and let
$A\in\M_n\otimes\A$. Thus, for each $\beta\in J$,
$\bigl(\id_{\M_n}\otimes\tilde{\de}_{\beta}P\bigr)(A)$ is a
continuous $\M_n$-valued function on $Y$. By the definition
\eq(normCXM) of the norm in $\M_n\otimes\B$ there is a linear
functional $\Phi_\beta$ on $\M_n$, of norm $\leq1$, such that
$$\norm{\bigl(\id_{\M_n}\otimes\tilde{\de}_{\beta}P
      \bigr)(A)}_{\M_n\otimes\B}
   =\norm{\bigl(\Phi_\beta\otimes\tilde{\de}_{\beta}P\bigr)(A)}_\B
      =\norm{\tilde{\de}_{\beta}P(\Phi_\beta\otimes\id_\A)(A)}_\B
\quad,$$
where we used that $\Phi_\beta:\M_n\to\Cx$, and
$\Cx\otimes\B\equiv\B$. We now introduce the function $\widehat
A\in\M\otimes\A$, defined as $\widehat A(\beta,x)=\Phi_\beta(A(x))$,
or $\widehat A(\beta,\cdot)=(\Phi_\beta\otimes\id_\A)(A)$.
Then the norm on the right hand side in the above equation is
smaller than
$$ \sup_{\beta'}\norm{\tilde{\de}_{\beta}P
           (\Phi_{\beta'}\otimes\id_\A)(A)}_\B
   =\norm{\bigl(\id_{\M}\otimes\tilde{\de}_{\beta}P
      \bigr)(\widehat A)}_{\M\otimes\B}
\quad.$$
Summing over $\beta$, and applying the given inequality for $\M$, we
find
$$\sum_{\beta\in J}\norm{\bigl(\id_{\M_n}\otimes
                 \tilde{\de}_{\beta}P \bigr)(A)}_{\M_n\otimes\B}
   \leq\sum_{\alpha\in I}\norm{\bigl(\id_{\M}
        \otimes\de_{\alpha}\bigr)(\widehat A)}_{\M\otimes\A}
\quad,$$
and the result (1) follows, because
$$\eqalign{
  \norm{\bigl(\id_{\M}\otimes\de_{\alpha}\bigr)
          (\widehat A)}_{\M\otimes\A}
   &= \sup_\beta \norm{\bigl(\Phi_\beta
         \otimes\de_{\alpha}\bigr)(A)}_{\A} \cr
   &\leq\sup_\Phi \norm{\bigl(\Phi
        \otimes\de_{\alpha}\bigr)(A)}_{\A}
    =\norm{\bigl(\id_{\M_n}
        \otimes\de_{\alpha}\bigr)(A)}_{\A}
\quad.}$$
To prove (2), note that, on a
two-dimensional algebra $\B$, all operators $\tilde\de_\beta$  with
$\tilde\de_\beta(\idty)=0$  are proportional, so we may replace the
definition of the oscillation norm by an equivalent one with
$\abs J=1$. Hence the claim follows from the observation that we
used (1) only with the now one-dimensional algebra $\M=\C(J)$.
\QED

However, in classical systems with more than two spin values per
site we may have strict inequality $\trnmcb{P}>\trnm{P}$. In the
following \Exa17/1x2Ising/ we even have $\trnmcb{P}>1>\trnm{P}$. Hence
$P$ contracts exponentially to its fixed point, and the same is true
for a non-interacting QCA with this one-site transition operator.
However, for a system with non-trivial propagation maps this
information is not sufficient, and only $\trnmcb{P}$ gives a bound
which is independent of the propagation.

For computing $\trnmcb{P}$ \Lem/decomposition/ may be helpful even in
the classical case, since then the norms $\cbnorm{G_{\beta\alpha}}$
can be replaced by ordinary norms.

We remark that with some of the modified definition of the
oscillation norms on the single site observable algebras described
at the beginning of Section~3.1, the stabilization can be avoided
altogether in the classical case \cite{Krist}. However, as already
remarked in that context, this would be in conflict with our
technique for proving the ergodicity criterion in the general
quantum case.

\example/1x2Ising/
We take $\A$ as a system of two Ising spins, with its standard
oscillation norm \eq(Isingosno). This norm can be written as
$$\def\sigmas{_{\sigma_1,\sigma_2}
}\eqalign{
  \trnm f&:= {1\over2}\max\sigmas
               \abs{f(-\sigma_1,\sigma_2)-f(\sigma_1,\sigma_2)}
            +{1\over2}\max\sigmas
               \abs{f(\sigma_1,-\sigma_2)-f(\sigma_1,\sigma_2) }  \cr
        &={1\over4}\abs{\sum\sigmas \sigma_1 f(\sigma_1,\sigma_2)}
         +{1\over4}\abs{\sum\sigmas \sigma_2 f(\sigma_1,\sigma_2)}
         +{1\over2}\abs{\sum\sigmas \sigma_1\sigma_2
                         f(\sigma_1,\sigma_2)}
\ ,}$$
where the variables $\sigma_i$ are $\pm1$. From the second form it
is obvious that the unit ball of $\trnm\cdot$ is the cartesian
product of $\Rl$ (corresponding to multiples of the identity), and an
octahedron in $\Rl^3$. We now consider a transition operator
$P:\A\to\A$, given by the matrix
$$ P={1\over12}\pmatrix{0& 1& 5& 6 \cr
                        0& 0& 2& 10\cr
                        0& 7& 5& 0 \cr
                        8& 0& 0& 4}
\quad,\deqno()$$
in a basis in which the components of $f$ are
$(f(++),f(+-),f(-+),f(--))$. The oscillation norm $\trnm{P}$ is
readily computed, by applying $P$ to the $6$ extreme points of the
octahedral unit sphere of $\trnm\cdot$, and computing the
oscillation norms of the images. The result is
$$  \trnm{P}={3\over4}
\quad.\deqno(1x2trnm)$$
We now couple the system to an additional Ising spin, denoted by
$\sigma_0$, and described in the algebra $\M=\Cx^2$. What we have to
estimate is the operator norm of \hfill\break
$(\id_\M\otimes P):\M\otimes\A\to\M\otimes\A$ with respect to the norm
$$\def\sigmas{_{\sigma_0,\sigma_1,\sigma_2}}  \eqalign{
 \trnm{f}_{\M\otimes\A}
   &:={1\over2}\max\sigmas
      \abs{f(\sigma_0,-\sigma_1,\sigma_2)-f(\sigma_0,\sigma_1,\sigma_2)}\cr
   &\ +{1\over2}\max\sigmas
      \abs{f(\sigma_0,\sigma_1,-\sigma_2)-f(\sigma_0,\sigma_1,\sigma_2)}
\quad.}$$
This norm is not characterized as easily as before. The unit ball
has two unbounded directions, and the compact convex set in the
remaining $6$ dimensions is bounded by $48$ hyperplanes. We did not
succeed in computing all the extreme points of this polytope, so we
have no explicit expression for $\trnm{\id_\M\otimes P}$.
However, any expression $\trnm{(\id_\M\otimes P)f}_{\M\otimes\A}$ with
$\trnm{f}_{\M\otimes\A}=1$, is a lower bound on
$\trnmcb{P}$. Taking for $f$ one of the $28$ extreme points of the
unit ball known to us, namely
$f=(f(+++),\ldots,f(---))=(2, 1, 1, 0, 0, -1, -1, 0)$, we find
$$ \trnmcb{P} \geq \trnm{\id_\M\otimes P} \geq {13\over12}
\quad.\deqno()$$
\endexample

\bgssection 4.2. Estimates in other norms

In this section, which is not needed later in this paper, we show
how an estimate of the form \eq(contractensor) fails, if we use some
criteria different from oscillation norms to define contractivity of
the factors. To us this is the main motivation for using oscillation
norms in the first place.

Since $P_x\idty=\idty$, ``contractivity'' has to be defined in a way
ignoring this known fixed point. A natural approach is to consider
contractivity in the {\bf quotient norm} of $\A/\Cx\idty$, i.e.
$$ \norm{A}':=\inf_{\lambda\in\Cx}\norm{A-\lambda\idty}
\quad.\deqno()$$
Similarly, for $P:\A\to\A$ we define
$$ \norm{P}':=\sup\Set\Big{\norm{PA}'\ }{\ \norm{A}'\leq1}
\quad,\deqno()$$
and call $P$ ``contractive'', if $\norm{P}'<1$. The following example
shows what kind of estimate we can expect for this norm of a tensor
product of transition operators.

\example/norm'/
We consider finite classical systems ($\A=\C(\Omega)$, $\Omega$ a
finite set), for which any transition operator is of the form
\eq(PCA1site)
$$ Pf(\omega)=\sum_\eta p(\omega,\eta)f(\eta)
\quad.$$
One easily checks that
$$ \norm{P}'={1\over2}\max_{\omega,\omega'} \sum_\eta
               \abs{p(\omega,\eta)-p(\omega',\eta)}
\quad.\deqno(norm')$$
{}From this formula, and the positivity and normalization conditions
for the tensor factors, one easily finds an estimate for tensor
products, namely
$$ \Norm\Big{\bigotimes_{x\in\Lambda}P_x}'
      \leq\sum_{x\in\Lambda}\norm{P_x}'
\quad.\deqno(norms'<)$$
Note that this estimate grows with the size of the region $\Lambda$,
and becomes completely useless for infinite regions. Hence the
question is whether this trivial estimate can be improved upon.

As a simple counterexample, consider an Ising spin system (\ie
$\Omega^x=\set{+,-}$ at each site $x$), and all factors $P_x\equiv
P_1$ equal. Let $\chi_+$ and $\chi_-$ denote the functions which are
$1$ on the points ``$+$'' and ``$-$'', respectively, and zero
otherwise. $P_1$ is characterized by the two probabilities
$$ p_{\pm}=p(\pm,+)= \bigl(P_1\chi_+\bigr)(\pm)
\quad.$$
{}From \eq(norm'), $\norm{P_1}'=\abs{p_+-p_-}$. For the products
$\chi_+^\Lambda=\bigotimes_{x\in\Lambda}\chi_+$ and
$P^\Lambda=\bigotimes_{x\in\Lambda}P_x$ over a finite set $\Lambda$
of $N$ sites we get $\norm{\chi_+^\Lambda}'=1/2$, and
$$\eqalign{
 \norm{P^\Lambda\chi_+^\Lambda}'
   &\geq {1\over2}\Abs\Big{
           \bigl(P^\Lambda\chi_+^\Lambda\bigr)(+\cdots+)
          -\bigl(P^\Lambda\chi_+^\Lambda\bigr)(-\cdots-)}  \cr
   &= {1\over2}\abs{p_+^N-p_-^N}
    \geq{1\over2}\abs{p_+-p_-}\ N \min\set{p_-,p_+}^{N-1} \cr
   &= N \min\set{p_-,p_+}^{N-1}\ \norm{P_1}'\
       \norm{\chi_+^\Lambda}'
\quad.}$$
Picking both $p_+$ and $p_-$ close to $1$, we see that
$\norm{P^\Lambda}'$ may come arbitrarily close to $N\norm{P_1}'$.
\endexample

Of course, the bound \eq(norms'<) cannot be improved upon in
quantum systems either, and the norm \eq(norm') remains useless in
infinite quantum systems, as well.

Another alternative to oscillation norms, which seems plausible at
first sight, is to use the observation that for norms of Hilbert
space contractions with a known fixed point the right hand side of
the analogue of \eq(norms'<) can be improved to a supremum. This
suggests the use of the {\bf Hilbert space norms}
$$ \norm{A}'_\omega
   =\bigl({\omega(A^*A) -\abs{\omega(A)}^2}\bigr)^{1/2}
\deqno(hnorm')$$
on $\A$, and its associated operator norms for some state $\omega$.
It follows from the complete positivity of transition operators that
$P$ is a contraction with respect to this norm, provided that
$\omega$ is invariant under $P$ (\ie $\omega\circ P=\omega$). This
may not seem like a severe restriction, since we know that any
transition operator admits an invariant state. We then define
$\norm{P}'_\omega$ as the best constant in the inequality
$\norm{P(A)}'_\omega\leq\ep\norm{A}'_\omega$.
The inequality
$$ \norm{P_1\otimes P_2}'_{\omega_1\otimes\omega_2}
      \leq\max_i \norm{P_i}'_{\omega_i}
$$
holds, and it seems that we achieved our goal of finding a quantity
that behaves well under composition.
However, there are several drawbacks.
First of all, the invariant states $\omega_i$ have to be explicitly
known in order to compute any norm. Secondly, the ergodicity
statement one gets from the inequality $\norm{P}'_\omega<1$ is
rather weak: it allows no conclusion about states of the infinite
system  which are singular with respect to $\omega$. Perhaps the
most severe restriction, however, is that the propagation operators
introducing interaction into the QCA setting by mixing different
cells also fail to be contractions with respect to this norm, so the
approach based on \eq(hnorm') seems to be limited to the trivial,
non-interacting case.

A similar criticism applies to the idea to use the {\bf spectral
radius} of $P$ as an operator on $\A/\Cx\idty$, denoted by
$\srp(P)$. Again, we have equality
$\srp(P_1\otimes P_2)=\max_i\srp(P_i)$, but we have no control over
this quantity for a product of two transition operators, which we
need to introduce interaction (see the proof of \Thm19/ergoQCA/). As
an elementary example consider, as in \Exa/1x2Ising/, a classical
system with two subcell types of one Ising spin each, with the
transition operator
$$ P_1=\pmatrix{0&1&0&0\cr0&1&0&0\cr0&1&0&0\cr
              0&0&1/2&1/2\cr}
\quad.$$
In a system consisting of two cells we choose the propagation map
$\dd(1){\cdot}$ for the first subcell type to be the identity, and
the map $\dd(2){\cdot}$ for the second subcell type to be the flip.
Then $P_1$ contracts exponentially with rate $1/2$ to its invariant
state, which is the pure state on the configuration $(+-)$. However,
the total transition operator $P$ has three invariant states.

Another conceivable alternative are the ``{\bf sup-oscillation
norms}'' introduced in \eq(strnm), taken now not only as a way to
define the oscillation norm in the subcells, but as a principle to
construct the total oscillation norm. Defining the operator norm
$\strnm{\cdot}_{\rm cb}$ in analogy to \Def/cbOscillationNorm/, it
is easy to show the analogue of \Thm/contractensor/. However, this
does not suffice to give an ergodicity criterion, since the estimate
\eq(osc1) fails for such norms, and consequently
$\strnm\cdot$-Cauchy sequences need not converge in $\A$. A simple
example demonstrating these claims is the sequence of averages of
Ising spins over an increasing sequence of regions. The
sup-oscillation norm of such averages goes to zero like the inverse
number of sites in the average, but, of course, the sequence of
averages is not convergent in the quasi-local algebra.

\bgsections 5. Applications to Cellular Automata
\bgssection 5.1. Ergodicity

In order to apply the results of the previous section to interacting
QCAs, we need tensorable oscillation norms on the algebras $\B^s$
belonging to each subcell type using, say operators
$\de^s_\alpha:\B^s\to\B^s$, $\alpha\in I^s$. Then by
\Prp/TensorableOscillation/ we have tensorable oscillation norms on
each $\A^x$, and, consequently, on the algebra $\A^\La$ of the whole
system. By \Prp/ergodicity/ ergodicity follows from the estimate
$\trnm P<1$ for the total transition operator $P$. Thus we arrive at
the following criterion.

\iproclaim/ergoQCA/Theorem.
Let a quantum cellular automaton be given according to \Def/QCA/,
and suppose that each $\B^s$ is equipped with a tensorable
oscillation norm. Then
$$ \trnmcb{P}=\trnmcb{P_1}
\quad.$$
Consequently, if $\trnmcb{P_1}<1$, the QCA is ergodic, \ie there is a
unique $P$-invariant state $\rho$ on $\A^\La$.
\eproclaim

\proof:
Let $\widetilde P=\bigotimes_{x\in\La} P_1$ be the infinite tensor
product of the operators $P_1$ acting in each $\A^x$ separately.
This is also the total transition operator of the QCA with the same
$P_1$, but each $\dd(s)\cdot$ equal to the identity. Consider also
the automorphism $D:\A^\La\to\A^\La$ which takes $\B^{(x,s)}$ into
$\B^{\dd(s)x,s}$. Then
$$ P=D\ \widetilde P
\quad.$$
According to \Thm/contractensor/,
$\trnmcb{\widetilde P}=\trnmcb{P_1}$. Moreover, since the
oscillation norm on $\B^{\dd(s)x,s}$ is defined by the same operators
$\de^s_\alpha$ as in $\B^{x,s}$, $D$ is a
$\trnm\cdot$-isometry, and $\trnmcb D=1$. Hence
$\trnmcb{P}=\trnmcb D\,\trnmcb{\widetilde P}=\trnmcb{P_1}$.
\QED

Note that similarly to the non-interacting case the criterion
$\trnmcb{P_1}<1$ is a condition of ``large disorder'', which is
satisfied as soon as $P_1$ is sufficiently close to a map of the
form $P_1(A)=\omega(A)\idty$ (compare \eq(disorder)). A remarkable
feature of this criterion is that it does not depend on the
propagation maps $\dd(s)\cdot$, which distinguish an interacting QCA
from a non-interacting one.

One might expect from \Lem/spectralP/ that, with a suitable choice of
oscillation norms, $\trnmcb{P_1}$ can be made equal to the
largest modulus of eigenvalues of $P_1$ apart from $1$. However, this
is not the case, since it is crucial for the proof of \Thm/ergoQCA/
that each oscillation operator $\de^s_\alpha$ acts in only one
subcell $\B^s$, and so the propagation automorphism $D$ becomes a
$\trnm\cdot$-isometry. Clearly, this property cannot be expected of
the eigenprojections of $P_1$. Still, the second largest modulus of
eigenvalues of $P_1$ is always a lower bound to $\trnmcb{P_1}$.

\bgssection 5.2. The classical case

Since in the usual definition of PCAs no subcell decomposition is
used, it is not obvious how the classical results
\cite{Lebowitz,MSa} can be subsumed under \Thm/ergoQCA/. In this
section we show how this can be done, pointing at the same time to a
possible generalization of \Def/QCA/.

The basic operator $P_1$  defining the one-site transition
probability of a PCA (see \eq(PCA1site)) maps the one-site algebra
$\A$ into the tensor product $\At=\A^{\otimes n}$, where $n$ is the
number of cells influencing the state in a single cell of the second
generation. $n$ is often finite, but we do not need this fact. We
will construct a QCA, whose one-site algebra is $\At$, with all
subcell algebras $\B^s$ isomorphic to $\A$. Note that all these
algebras are now commutative, so the ``Q'' in QCA only refers to the
fulfillment of \Def/QCA/.

The total PCA transition operator $P$ (see \eq(indup)) can be
decomposed into three factors: the first is simply the infinite
tensor product $P_1^\La$ of the operator $P_1$, mapping $\A^\La$ to
$\At^\La$. The information about the cell $y$ to which a subcell
$(x,s)$ in the latter algebra belongs in the next time step is
encoded in propagation maps
$\dd(s){\cdot}:\La\to\La$ as before, \ie $y=\dd(s)x$.
This defines an automorphism $D$ of $\At^\La$ as in the proof of
\Thm/ergoQCA/. The final step is the sitewise application of the
multiplication map $\diag:\At\to\A$, defined by
$$ \diag(f_1\otimes\cdots \otimes f_n)=\prod_{i=1}^n f_i
\quad.\deqno(diagmul)$$
Hence we get the factorization
$$ P=\diag^\La\ D\ P_1^\La
\quad,\deqno(PCAfact)$$
where $\diag^\La$ denotes the tensor product of the copies of $\diag$
acting at each site.

It is precisely the use of the multiplication map $\diag$, that is
the specifically classical element in this construction. $\diag$ is
also called the $n\th$ order {\it diagonal} of the algebra $\A$
because, writing $\A\cong\C(\Omega)$, and
$\A^{\otimes n}=\C(\Omega^n)$, as we may for an abelian algebra, we
have
$$ \diag f(x) =f(x,x,\ldots,x)
\quad.\deqno(diag)$$
Clearly, $\diag$ is a *-homomorphism, and
$\diag(\idty\otimes\cdots\otimes f\otimes\cdots\otimes\idty)=f$, for
all positions of the factor $f$ in the tensor product. The existence
of the diagonal characterizes abelian algebras: if a homomorphism
$\diag$ of this description exists in a C*-algebra $\A$, the
commutativity of the tensor multiplication implies the commutativity
of multiplication. The adjoint of the diagonal map is a ``state
duplication map'', producing, from a state on $\A$, $n$ copies of
$\A$ in the same state. Its non-existence in the quantum case is the
basis of ``quantum cryptography'' (see \cite{qcrypt} and references
cited there). Of course, we can formally define $\diag$ by
\eq(diagmul), even in the non-commutative case. However, if
$\A=\M_d$, $n=2$, and $\Phi$ is the unitary permutation operator
exchanging the two factors, $\norm{\diag(\Phi)}=n$, \ie
$\norm{\diag}\geq n$. Hence, on an infinite dimensional algebra,
$\diag$ is typically unbounded. But even in the finite dimensional
case, $\norm{\diag}>1$ makes the definition of the infinite tensor
product $\diag^\La$ in \eq(PCAfact) impossible.

It is now easy to modify \eq(PCAfact) so that we get a QCA in the
sense of \Def/QCA/. Its transition operator is
$$ \widetilde P
      :=  D\ P_1^\La\ \diag^\La
\quad.\deqno(QCAfact)$$
One easily verifies that
this is the QCA with one-site transition operator
$$\eqalign{
  \widetilde P_1&:\At\to\At \cr
  \widetilde P_1&=P_1 \diag
\quad.}$$
With $J:\A\to\At$, defined as $Jf=f\otimes\idty^{\otimes(n-1)}$, we
have $\diag J=\id_\A$, and hence
$$ P^N= \diag^\La\ {\widetilde P}^N\ J^\La
\quad,\deqno(dilation)$$
for every power $N\geq0$, \ie the PCA can be recovered completely
from the QCA picture.

For estimating the contraction rates of these operators we need the
following Lemma.

\iproclaim/trnm:diag/Lemma.
Let $\A=\C(\Omega)$ be a finite dimensional abelian C*-algebra with
an oscillation norm defined by operators $\de_{\alpha}$ of the form
$$ \de_{\alpha}f(\sigma)
      =c_\alpha\Bigl( f(a_\alpha(\sigma)) -f(\sigma)\Bigr)
\quad,$$
where $c_\alpha\in\Rl$, and $a_\alpha:\Omega\to\Omega$.
Then, for every $n$, the $n\th$ order diagonal $\diag$ satisfies the
estimate $\trnmcb{\diag}\leq1$.
\eproclaim

\proof:
$\M_d\otimes\A$ can be identified with the algebra of
$\M_d$-valued functions on $\Omega$, equipped with the norm
$\norm{g}=\sup_\sigma\norm{g(\sigma)}$. Then, for
$f\in\M_d\otimes\A^{\otimes n}$ and $\sigma'=a_\alpha(\sigma)$, the
expression
$$\def\ub#1#2{\underbrace{\sigma{#1},\ldots,\sigma{#1}}_{#2\ \rm times}}
  F_k=f(\ub'k,\ \ub{}{n-k})-f(\ub'{k-1},\ \ub{}{n-k+1})
$$
is bounded in the norm of $\M_d$ by
$\abs{c_\alpha^{-1}}\Norm\big{
           (\id_{\M_d}\otimes\de_\alpha^{(k)})f}$. Hence
$$\eqalign{
    \norm{\bigl((\id_{\M_d}\otimes\de_\alpha\diag) f\bigr)(\sigma)}
       &=\abs{c_\alpha}\
           \norm{f(\sigma',\ldots,\sigma')
                -f(\sigma,\ldots,\sigma)} \cr
       &=\abs{c_\alpha}\ \Norm\Big{\sum_{k=1}^{n-1} F_k}  \cr
       &\leq \sum_{k=1}^{n-1}
           \Norm\Big{(\id_{\M_d}\otimes\de_\alpha^{(k)})f}
\quad.}$$
Hence, after taking the supremum over $\sigma$, summing over
$\alpha$, and using the definition \eq(tensorTrnm) of the canonical
oscillation norm on $\A^{\otimes n}$, we get \eq(trnmcb) with
$\ep=1$. The constant cannot be better than $1$, because
$\diag J=\id_\A$, and, obviously $\trnm{J f}=\trnm{f}$ for all $f$.
\QED

Hence, from the factorization \eq(dilation)
we get
$\trnmcb{P}\leq \trnmcb{\diag^\La}\trnmcb{\widetilde P}
                \trnmcb{J^\La}
           \leq \trnmcb{\widetilde P}$.
{}From \eq(QCAfact) and because $D$ is an oscillation norm isometry,
$\trnmcb{\widetilde P}\leq \trnmcb{P_1^\La}$.
Because $P_1^\La=D^{-1}\widetilde PJ^\La$, the last inequality is
actually an equality. By \Thm/contractensor/,
$\trnmcb{P_1^\La}=\trnmcb{P_1}$. Summing up these estimates, we have
$$ \trnmcb{P}\leq \trnmcb{\widetilde P}=\trnmcb{P_1}
\quad.$$
This is exactly the bound given in \cite{Lebowitz}. When comparing
these results, however, note that \Prp/ergodicity/ gives
$\norm{P^N(A)-\rho(A)\idty}\leq2\ep^n\trnm A$, without the
superfluous factor $(1-\ep)^{-1}$, which is present in
\cite{Lebowitz}.

Finally, we wish to point out that \eq(QCAfact) also points to a
possible generalization of the notion of QCA, which does not use
subcell decompositions: we only have to replace $\diag$ by some
completely positive unital operator from $\A^{\otimes n}\to\A$. For
such systems our method for obtaining oscillation norm estimates
would apply unchanged, but they would no longer satisfy the
condition of commuting ranges (see Section 2). In this sense the
cells would no longer be ``independently updated''.

\bgssection 5.3. Decay of correlations in the invariant state

We assume now that the ergodicity criterion $\trnmcb{P_1}<1$ holds.
What can be said about the unique invariant state $\rho$ to which
$P$ contracts? It is clear that in the non-interacting case
(\ie $\dd(s)\cdot=\id$), but also if $P_1(A)=\omega(A)\idty$ (\ie
$\trnmcb{P_1}=0$), $\rho$ will be a product state. Therefore, it is
reasonable to expect that if $\trnmcb{P_1}$ is small, we should
obtain a state with good clustering properties. Moreover, in
contrast to the ergodicity criterion, the propagation maps
$\dd(s)\cdot$ should enter the estimate for the correlation
functions.

We will first describe the relevant geometric properties of the
$\dd(s)\cdot$. For $\Lambda\subset\La$, we will set
$$ d_S(\Lambda)=\Set\big{\dd(s)x}{s\in\Subc,\ x\in\Lambda}
                \subset\La
\quad.$$
This is the set of cells to which the interaction can spread from
some site in $\Lambda$ in one step. Similarly, we define
$d_S^n(\Lambda)$ as the $n\th$ iterate of $d_S$.
For $\Lambda_1,\Lambda_2\subset\La$ we define the {\it correlation
distance} as
$$ c(\Lambda_1,\Lambda_2)
    =\max\Set\big{n\in\Nl}{d_S^n(\Lambda_1)\cap
            d_S^n(\Lambda_2)=\emptyset}
\quad,$$
\ie as the last time step under which the two regions remain
independent. When the maps $\dd(s)\cdot$ are translations, it is
clear that for large separation parameters $r$, the correlation
distance  $c(\Lambda_1, \Lambda_2+r\vec e)$ will asymptotically be
proportional to $r$, but with a constant depending on the direction
$\vec e$. In this sense the following Proposition gives exponential
clustering with a rate depending both on the direction, and on
$\trnmcb{P_1}$.

\iproclaim/CorrelationDecay/Proposition.
Let a quantum cellular automaton be given according to \Def/QCA/,
and suppose that, with respect to some choice of tensorable
oscillation norms on each $\B^s$, the ergodicity criterion
$\trnmcb{P_1}<1$ is satisfied. Let $\rho$ denote the unique state
such that $\rho\circ P=\rho$.
Let
$$
  A_1 \in\A^{\Lambda_1} \,=\, \bigotimes_{x\in\Lambda_1} \A^x, \quad
  A_2 \in\A^{\Lambda_2} \,=\, \bigotimes_{x\in\Lambda_2} \A^x
\quad, $$
where $\Lambda_1\cap\Lambda_2=\emptyset$ are disjoint finite subsets
of $\La$.
Then
\item{(1)}
$P^k(A_1)\subset\A^{d_S^k(\Lambda_1)}$ for all $k\in\Nl$,
\item{(2)} $P^k\bigl(A_1\otimes A_2\bigr)= P^k(A_1)\otimes P^k(A_2)$,
for all $k\leq c(\Lambda_1,\Lambda_2)$.
\item{(3)}
$\abs{\rho(A_1\otimes A_2) - \rho(A_1)\rho(A_2)} \le
  2\,\bigl( \trnmcb{P_1}\bigr)^{c(\Lambda_1,\Lambda_2)} \,
     \Bigl(\trnm{A_1}\,\norm{A_2}
             + \norm{A_1}\, \trnm{A_2}\Bigr)$.
\eproclaim

\proof:
The first two statements are obvious from \Def/QCA/ for $k=1$, and
follow for other $k$ by induction.
Then using \Prp/ergodicity/ and  \Thm/ergoQCA/, we get, for all
$k\leq c(\Lambda_1,\Lambda_2)$, the estimate
$$\eqalign{ &\mkern-20mu
 \abs{\rho(A_1\otimes A_2) - \rho(A_1)\rho(A_2)}
     =\norm{\rho(A_1\otimes A_2)\idty - \rho(A_1)\rho(A_2)\idty} \cr
    &\le \norm{\rho(A_1\otimes A_2)\idty - P^k(A_1\otimes A_2)} +
          \norm{P^k(A_1\otimes A_2) - P^k(A_1)\otimes P^k(A_2)} \cr
    &\quad + \norm{P^k(A_1) - \rho(A_1)\idty}\,\norm{P^k(A_2)} +
          \abs{\rho(A_1)}\,\norm{P^k(A_2) - \rho(A_2)\idty} \cr
    &\le \bigl (\trnmcb{P_1}\bigr)^k\trnm{A_1\otimes A_2} +
          \bigl(\trnmcb{P_1}\bigr)^k \norm{A_2} \, \trnm{A_1} +
          \bigl(\trnmcb{P_1}\bigr)^k \norm{A_1}\,\trnm{A_2},
}$$
where at the last step we used that $\norm{P^k(A_2)}\leq\norm{A_2}$.
The result then follows from equation \eq(LocalOsc).
\QED

\bgsection . Appendix: Complete boundedness

To motivate the necessity of considering complete positivity and
complete boundedness of operators on non-commutative C*-algebras we
consider a standard example: the operator $P:\M_n\to\M_n$ of
transposition on the algebra of $n\times n$-matrices. This preserves
positivity and the identity element, and therefore seems to be a
candidate for a transition operator. However, positivity and the
norm bound $\norm{P}\leq1$ both get lost if we consider the system
as a subsystem of a larger one with observable algebra, say
$\M_n\otimes\M_n$. Then $(\id_{\M_n}\otimes P)$ takes the unitary
flip operator $\Phi=\sum_{ij}\abs{ij\rangle\langle ji}$ into $n$ times
the one-dimensional projection
$p=(1/n)\sum_{ij}\abs{ii\rangle\langle jj}$. Hence
$\norm{\id_{\M_n}\otimes P}\geq n$, and $P(\idty-\Phi)=\idty-n p$ is
not positive, although $(\idty-\Phi)$ is. Clearly, ${\id_{\M_n}\otimes
P}$ is no longer a transition operator, although there is no
interaction with the ``innocent bystander'' system described in
$\M_n$.

In order to exclude such phenomena one defines a linear operator
$P:\A\to\B$ between C*-algebras to be {\it completely
positive} \cite{TAK,PAUc}, if $\id_{\M_n}\otimes P$ is positive for
all $n$ or, equivalently \cite{TAK}, if for any choice of $n$-tuples
$a_1,\ldots,a_n\in\A$, and $b_1,\ldots,b_n\in\B$, the operator
$\sum_{ij}b_i^*P(a_i^*a_j)b_j$ is positive. $P$ is said to be {\it
completely bounded}, if $\norm{\id_{\M_n}\otimes P}$ is bounded by a
constant independent of $n$. For such operators we define the {\it
completely bounded norm} as
$$ \cbnorm{P}:=\sup_{n\in\Nl}\norm{\id_{\M_n}\otimes P}
\quad.$$
If $\cbnorm{P}\leq1$, $P$ is called a {\it complete contraction}.
The appearance of the matrix algebras $\M_n$ in these definitions is
solely a matter of convenience: these definitions imply the
corresponding statements with $\M_n$ replaced by an arbitrary
C*-algebra $\M$ (see the proof of \Lem/allM/ for a very similar
argument).
For checking complete positivity or boundedness it is often
useful to consider $\M_n\otimes\A$ as the *-algebra of $n\times
n$-matrices with entries in $\A$. When $\B$ is a finite
dimensional algebra containing as direct summands at most the
$k\times k$-matrices, it suffices to verify complete positivity, or
to compute $\cbnorm P$, in $\M_n\otimes\A$ with $n=k$ \cite{SMI}. In
particular, all operators between finite dimensional C*-algebras are
completely bounded.

Basic examples of completely positive maps are *-homomorphisms, maps
of the form $A\mapsto V^*AV$, and all positive maps with either $\A$
or $\B$ abelian, which includes all states. Completely positive
operators are completely bounded, and when $\idty\in\A$, and $P$ is
completely positive, we have $\norm{P(\idty)}=\norm{P}=\cbnorm{P}$
\cite{PAUc}. The fundamental structure theorem for completely
positive maps is the Stinespring Dilation Theorem \cite{STI},
stating that every completely positive $P:\A\to\B(\H)$ can be
decomposed in an essentially unique way into $P(A)=V^*\pi(A)V$,
where $\pi:\A\to\B(\K)$ is a *-representation of $\A$ on a Hilbert
space $\K$, and $V:\H\to\K$ is a bounded operator.

Basic examples of complete contractions are differences $P=P_+-P_-$
of completely positive maps with $\norm{P_++P_-}\leq1$, and
multiplication operators $A\mapsto MA$ where $\norm{M}\leq1$.
Results analogous to the Stinespring dilation are also available for
completely bounded maps. However, the uniqueness is typically lost.
Thus any complete contraction $P:\A\to\B(\H)$ with $\A\ni\idty$ can
be decomposed as $P(A)=V_1^*\pi(A)V_2$, with $\pi$ a
*-representation of $\A$, and $V_1$ and $V_2$ isometries.
Essentially the same statement is that every complete contraction
$P:\A\to\B(\H)$ can be realized as the off-diagonal corner of a
completely positive map \cite{PAUc}, \ie there is a unit
preserving completely positive map
$\widetilde P:\M_2\otimes\A\to\M_2\otimes\B(\H)$, such that
$$ \widetilde P:\left( \pmatrix{0 & A \cr 0 & 0} \right)
           = \pmatrix{0 & P(A) \cr 0&0}
\quad.$$
Every completely bounded operator $P:\A\to\B(\H)$ is a linear
combination of completely positive ones.
If, moreover, $P$ is hermitian (\ie $P(A^*)=P(A)^*$), one can find a
completely positive $P_+$ with $\cbnorm{P_+}=\cbnorm{P}$ such that
$P_+\pm P$ are completely positive \cite{WIT,PAUc}. The same
statement holds when $\B(\H)$ is replaced by an arbitrary injective
C*-algebra \cite{WIT}, but fails in general. Since finite
dimensional algebras are injective, this covers the applications of
this result in connection with \Lem/decomposition/.

\let\REF\doref
\Acknow
We gratefully acknowledge stimulating discussions with Mark Fannes,
Aernout van Enter, Burkhard K\"ummerer, and Christian Maes.
S.R. was supported by a scholarship from the DFG (Bonn) under the
``Graduiertenkolleg'' programme, and by the scholarship
KUL-OT/92-09 (Leuven).

\REF Bia Biafore \Jref
    M. Biafore
    "Cellular automata for nanometer-scale computation"
    Physica D @70(1994) 415--433

\REF BR BraRo      \Bref
    O. Bratteli, D.W. Robinson
     "Operator algebras and quantum statistical mechanics"
     2 volumes, Springer Verlag, Berlin, Heidelberg, New York
     1979 and 1981

\REF DiV Vincenzo \Jref
    D.P. DiVincenzo
    "Two-bit gates are universal for quantum computation"
    Phys.Rev. A @51(1995) 1015--1022

\REF DS Dunfo \Bref
    N. Dunford, J.T. Schwartz
    "Linear operators, I"
    Wiley\&Sons, New York 1957

\REF EHPP qcrypt \Jref
    A.K. Ekert, B. Huttner, G.M. Palma, A. Peres
    "Eavesdropping on quantum-cryptographical systems"
    Phys.Rev. A @50(1994)1047--1056

\REF FNW1 FCS      \Jref
    M. Fannes, B. Nachtergaele, R.F. Werner
    "Finitely correlated states of quantum spin chains"
    Commun.Math.Phys. @144(1992) 443--490

\REF FNW2 FCP      \Jref
    M. Fannes, B. Nachtergaele, R.F. Werner
    "Finitely correlated pure states"
    J.Funct.Anal. @120(1994) 511--534

\REF FGSS Zei2 \Jref
    S. Fussy, G. Gr\"ossing, H. Schwabl, A. Scrinzi
    "Nonlinear computation in quantum cellular automata"
    Phys.Rev. A @48(1993) 3470--3477

\REF GD GD \Jref
  A. Georges, P. Le Doussal
  "From equilibrium spin models to probabilistic cellular automata"
  J.Stat.Phys. @54(1989) 1011--1064

\REF GZ  Zei1 \Jref
    G. Gr\"ossing, A. Zeilinger
    "Quantum cellular automata"
    Complex Syst. @2(1988) 197--208, and 611--623

\REF LMS Lebowitz \Jref
    J.L. Lebowitz, C. Maes, E.R. Speer
    "Statistical mechanics of probabilistic cellular automata"
    J.Stat.Phys. @59(1990) 117--170

\REF LTPH Lent \Jref
    C. Lent, P.D. Tougaw, W. Porod, G.H. Bernstein
    "Quantum cellular automata"
    Nanotechnology @4(1993) 49--57

\REF Lig Liggett \Bref
    T.M. Liggett
    "Interacting particle systems"
    Springer-Verlag, New York 1985

\REF Mae Krist \Jref
    C. Maes
    "Coupling interacting particle systems"
    Rev.Math.Phys. @5(1993) 457--475

\REF MS1 MSa \Jref
    C. Maes, S.B. Shlosman
    "Ergodicity of probabilistic cellular automata: a constructive
    criterion"
    Comm.Math.Phys. @135(1991) 233--251

\REF MS2 MSc \Jref
     C. Maes, S. Shlosman
     "When is an interacting particle system ergodic?"
     Commun.Math.Phys. @151(1993) 447--466

\REF Mai Mainieri \Gref
    R. Mainieri
    "Design constraints for nanometer scale quantum computers"
    Preprint Los Alamos LAUR93-4333,
    archived as {\tt cond-mat/9410109}

\REF MZ Boguslav \Gref
    A.W. Majewski, B. Zegarlinski
    "Quantum stochastic dynamics I:
     spin systems on a lattice"
    Preprint, Imperial College London, March 1995

\REF Ma1 MATA   \Jref
  T. Matsui
  "On Markov semigroups of UHF algebras"
  Rev.Math.Phys. @5(1993) 587--600

\REF Ma2 MATB   \Jref
    T. Matsui
    "Purification and uniqueness of quantum Gibbs states"
     Commun.Math.Phys. @162(1994) 321--332

\REF Ma3 MATC \Gref
    T. Matsui
    "Interacting particle systems on non-commutative spaces"
    \inPr M. Fannes, C. Maes, A. Verbeure
    "On three levels; micro-, meso-, and macro-approaches in physics"
    Plenum Press, New York 1994

\REF Ma4 MATD \Gref
    T. Matsui
    "Quantum statistical mechanics and Feller semigroup"
    Preprint Tokyo Metropolitan University 1995

\REF Pau PAUc \Bref
  V.I. Paulsen
  "Completely bounded maps and dilations"
  Pitman Research Notes in Mathematics, Longmans, London 1986

\REF Ric PHD \Gref
    S. Richter
    "Construction of states on two-dimensional lattices
     and quantum cellular automata"
     PhD Thesis, Osnabr\"uck 1994

\REF Smi SMI    \Jref
     R.R. Smith
     "Completely bounded maps between C*-algebras"
     J.London.Math.Soc. @27(1983) 157--166

\REF Sti STI \Jref
    W.F. Stinespring
    "Positive functions on C*-algebras"
    Proc.Amer.Math.Soc. @6(1955) 211--216

\REF Tak TAK      \Bref
    M. Takesaki "Theory of operator algebras I"
    Springer Verlag, Berlin, Heidelberg, New York 1979

\REF Wit WIT \Jref
  G. Wittstock
  "Ein operatorwertiger Hahn-Banach Satz"
  J.Funct.Anal. @40(1981) 127--150

\bye